\documentclass[preprint,notoc]{JHEP3}
\usepackage{epsfig}
\usepackage{cite}

\def\to{\rightarrow}

\def\bi{\begin{itemize}}
\def\ei{\end{itemize}}

\def\c1p{C1^\prime}
\def\ta{\tilde a}
\def\tG{\widetilde G}

\def\ta{\tilde a}

\def\tg{\tilde g}

\def\tB{\widetilde B}

\def\tz{\widetilde Z}

\def\alt{\lesssim}
\def\agt{\gtrsim}
\def\to{\rightarrow}
\def\To{\Rightarrow}
\def\be{\begin{equation}}  
\def\ee{\end{equation}}  
\def\bea{\begin{eqnarray}}  
\def\eea{\end{eqnarray}}  

\newcommand\njp[3]{{\it New\ J.\ Phys.\ }{\bf #1} (#2) #3}

\newcommand\annp[3]{{\it Annals\ Phys.\ }{\bf #1} (#2) #3}
\newcommand\sjp[3]{{\it Sov.\ J.\ Nucl.\ }{\bf #1} (#2) #3}

\title{Some necessary conditions for allowing the PQ scale\\
 as high as $M_{GUT}$ in SUSY models\\
 with an axino or neutralino LSP}

\author{Howard Baer$^{a}$ and Andre Lessa$^{a}$\\
$^a$Dept.\ of Physics and Astronomy, University of Oklahoma, Norman, OK 73019, USA\\
E-mail: \email{baer@nhn.ou.edu}, \email{lessa.a.p@gmail.com}}

\abstract{We examine some conditions which are needed to allow the Peccei-Quinn scale ($f_a$) 
as high as the SUSY GUT scale ($M_{GUT} \sim 10^{16}~GeV)$ in the context of the PQMSSM with 
either an axino or a neutralino LSP. 
The main problem in non-SUSY models with $f_a\sim M_{GUT}$ is the generation of an overabundance of
axion dark matter ($\sim f_a^{7/6}$) due to vacuum misalignment. 
We show that once all the components of the axion supermultiplet
are included, the upper limit on $f_a$ can be evaded due to large entropy injection from saxion decays.
This large entropy injection also dilutes all other quasi-stable relic densities, naturally evading
the BBN constraints and solving the gravitino problem.
We find that $f_a\sim M_{GUT}$ can be allowed by relic density/BBN 
constraints provided that the saxion mass $m_s\agt 50$ TeV, the initial saxion field value is
of order of the PQ scale and the initial axion mis-alignment angle
$\theta_i\alt 0.05$. These restrictions can be considerably loosened for  $f_a\sim 10^{15}$ GeV.
The allowed range for the re-heat temperature ($T_R$) is strongly dependent on the
nature of the LSP. For the axino LSP, $T_R \gtrsim 10^8$~GeV, while for neutralino LSP any value is allowed.
In the latter case, $f_a\sim M_{GUT}$ can be more easily accommodated 
than in the axino LSP scenario.
For $f_a\sim M_{GUT}$ in SUSY models, the dark matter abundance should be dominated by axions, 
albeit with mass $m_a\sim 10^{-10}$ eV, 
far below the region currently probed by experiment.
}

\begin{document}

\section{Introduction}
\label{sec:intro}

The strong $CP$ problem remains one of the
central puzzles of QCD which evades explanation within the context of
the Standard Model. The crux of the problem is that an additional
$CP$ violating term in the QCD Lagrangian of the form\footnote{Here $G_{\mu\nu}^a$ 
is the gluon field strength tensor and $\tilde{G}^{a\mu\nu}$ its dual.} 
$\bar{\theta}g_s^2/32\pi^2 G^{\mu\nu}_A \widetilde{G}_{A\mu\nu}$ 
ought to be present as a result of the t'Hooft resolution of the $U(1)_A$
problem via instantons and the $\theta$ vacuum of QCD~\cite{thooft}. 
Here, $\bar{\theta}\equiv \theta +arg\ det{\cal M}$ actually consists of two terms:
one from QCD and one from the electroweak quark mass matrix.
The experimental limits on the neutron electric dipole moment
however constrain $|\bar{\theta}|\alt 10^{-10}$~\cite{nedm}. 
Explaining why the sum of these two terms should be so small is the essence of the 
strong $CP$ problem\cite{scp_review}.

An extremely compelling solution proposed by Peccei and Quinn~\cite{pq} is
to hypothesize an additional global $U(1)_{\rm PQ}$ symmetry, which is broken at
some mass scale $f_a \gtrsim 10^9$~GeV. A consequence of the broken PQ symmetry is the existence of a 
pseudo-Goldstone boson field: the axion $a(x)$~\cite{ww}.
The low energy Lagrangian then includes the interaction term
\be
{\cal L}\ni \frac{\alpha_s}{8\pi}\frac{a(x)}{f_a/N} G_{\mu\nu}^a \tilde{G}^{a\mu\nu} \,, \label{L_ax}
\ee
where $\alpha_s = g_s^2/4\pi$, $g_s$ is the QCD coupling constant and
$N$ is the model-dependent color anomaly factor. (From here on we  assume $N=1$, but
all our results can be extended to any $N$ value with the replacement $f_a \to f_a/N$.)
Since $a(x)$ is dynamical, the entire $CP$-violating term settles to its
minimum at zero, thus resolving the strong $CP$ problem. 
A consequence of this very elegant mechanism is that
a physical axion field should exist, with axion excitations of mass\cite{bardeen}
\be
m_a\simeq 6\ {\rm eV}\ \frac{10^6\ {\rm~GeV}}{f_a} \, .
\label{eq:axmass}
\ee

Due to its tiny mass, the interactions of the axion field need to be
strongly suppressed-- $f_a\agt 10^9$ GeV--
otherwise they would have a strong impact on low energy physics and astrophysics\cite{astro},
such as too rapid cooling of stars and supernovae. 
On the other hand, under some model assumptions (see Sec.~\ref{sec:DMconst} below), 
the axion relic density $\Omega_{a} h^2 < 0.1123$ bound requires $f_a \lesssim 10^{12}$~GeV. 
As a consequence a new physics scale much larger than
the electroweak scale ($\sim m_W$) has to be introduced into the Standard Model\cite{ksvz,dfsz}. 
This results in large radiative corrections to the Higgs mass, which then requires a large amount of fine-tuning
to stabilize the EW scale. This leads to the well known hierarchy problem.

So far, one of the most compelling solutions to naturally stabilize the EW scale is supersymmetry
(SUSY), which reduces quadratic divergences to merely logarithmic, and ameliorates the fine-tuning problem\cite{wss}.
The supersymmetric version of the Standard Model (MSSM) also has other compelling features, such as  viable
Dark Matter (DM) candidate(s) and unification of the gauge coupling constants at $M_{GUT} \simeq 2\times10^{16}$~GeV.
In order to accomodate the PQ solution in the MSSM, it is necessary to postulate the existence of new
superfields carrying PQ charge. Although this extension of the MSSM can be realized in a number of ways,
we will call the resulting weak-scale effective theory the PQMSSM.

It has been noticed early on\cite{Kim84} that the PQ scale $f_a$ falls within the desired range for the SUSY
breaking scale ($m_{SUSY}$) in gravity-mediated SUSY breaking models:
\be
m_{3/2} \sim m_{SUSY}^2/M_P \sim 1\; {\rm TeV} \Rightarrow m_{SUSY} \sim 10^{11}\; {\rm~GeV}\; \sim f_a ,
\ee
where $m_{3/2}$ is the gravitino mass and $M_P$ is the reduced Planck mass. 
As a result, several models have been proposed\cite{Kim84,susypq,chun2011} to connect the SUSY and PQ breaking scales. 
This can be achieved if the axion supermultiplet has tree level interactions with the 
hidden sector responsible for breaking SUSY.

However, once a grand unified theory is assumed, the $U(1)_{PQ}$ symmetry can appear as an accidental global
symmetry of the theory, as naturally occurs in several SUSY GUTS\cite{nr}.
In this case, $f_a$ will naturally be of order $M_{GUT}$ and will strongly violate its $10^{12}$~GeV upper limit.
It is possible to protect $f_a$ from obtaining $M_{GUT}$ contributions, either by breaking the PQ symmetry
at a lower scale or artificially suppressing the vacuum expectation value of the axion supermultiplet. Nonetheless
such mechanisms always require the introduction of new superfields or fine-tuned parameters only for this purpose.

String theory has emerged as an attractive ultraviolet complete theory which can easily incorporate axion-like
fields as elements of anti-symmetric tensors\cite{o32}. Many of the would-be axions become massive, while the 
remaining light fields obey a global PQ symmetry. A survey of a variety of string models\cite{witten,dfkw} 
indicates that while PQ symmetry is easy to generate in string theory, the associated PQ scale
tends to occur at or near the GUT scale rather than some much lower intermediate scale. This is in 
apparent conflict with the simple limits on $f_a$ from overproduction of dark matter as discussed above.

One solution to the apparent conflict which allows for $f_a\sim M_{GUT}$ is to invoke a tiny initial 
axion mis-alignment angle $\theta_i\sim 0.003$. In this case, one must accept a highly fine-tuned
initial parameter which might emerge anthropically.

An alternative solution was proposed in one of the original papers calculating the cosmic abundance
of relic axions\cite{df}: perhaps additional massive fields are present in the theory, whose late decays
can inject substantial entropy into the universe at times after axion oscillations begin, but before BBN starts.
In Ref. \cite{df}, it was proposed that the gravitino might play such a role.
Several subsequent works have also explored the issue of dilution of (quasi)-stable relics 
via entropy injection\cite{entropy,kim1991,kmy,cck,gutfa,fox,bobkov,hasen}.

Therefore it is of interest to investigate under which conditions the PQ scale can be extended
to the GUT scale, while avoiding the known experimental constraints. While this possibility has
been suggested before\cite{kmy,gutfa}, we wish to explore this case using detailed 
calculations of particle production rates coupled with recent constraints arising from Big Bang nucleosynthesis
(BBN). 
Here, we investigate the implications of $f_a \sim M_{GUT}$ from a phenomenological point of view.
In order to keep our conclusions as general as possible we will avoid choosing a specific GUT
theory or PQMSSM model whenever possible.

In Sec.~\ref{sec:pheno}, we will present the general features of PQMSSM cosmology used
in our analysis and review several well known results for this model. We will
then discuss the $f_a \sim M_{GUT}$ scenario in Sec.~\ref{sec:bigfa} and the case of a light
axino LSP in Sec. \ref{sec:axino}. We will present the main difficulties associated with large $f_a$ values
and show how they can be avoided in the PQMSSM framework. 
In Sec. \ref{sec:az1}, we discuss the scenario where $m_{\ta}\sim m_{3/2}$, where the
neutralino is assumed to be LSP. In several respects, this scenario is more appealing than the light axino LSP 
case.
Section~\ref{sec:conclude} summarizes our main results and discusses the implications of unifying the PQ
and GUT scales. The appendices contain explicit formulae for evaluating the axion and neutralino
relic abundances in radiation-, matter- and decaying-particle- dominated universes.

\section{PQMSSM Phenomenology}
\label{sec:pheno}

In order to implement the PQ mechanism in supersymmetric theories, 
PQ charges have to be assigned to the MSSM fields and new PQ superfields must be introduced.
The axion superfield is obtained from linear combinations of other elementary (non-MSSM) fields 
and is a singlet under the MSSM gauge group. 
Even though the full field content of the PQMSSM is highly model dependent,
it must contain an axion supermultiplet composed of a complex scalar field ($\phi$) and a Majorana
fermion ($\tilde{a}$). The complex scalar field is usually divided into its axion ($a$) and
saxion ($s$) components:
\[ \phi = \frac{s + i a}{\sqrt{2}} \]
and the fermionic component is named {\it axino}.

Since the axion field is the $U(1)_{PQ}$ pseudo-Goldstone boson,
 it is massless, except for anomalous corrections coming from the QCD chiral
anomaly. For temperatures well above $\Lambda$ (the QCD chiral breaking scale),
the axion is essentially massless, while for $T\ll \Lambda$ the QCD chiral anomaly
induces a non-zero mass for the axion field. The temperature dependent axion mass is
given by\cite{ma,vg1}:
\be
m_a(T) = \left\{ \begin{array}{ll}
m_a^0 & \mbox{, if $T<\Lambda$} \nonumber \\
m_a^0 \times 0.018 \left(\frac{\Lambda}{T}\right)^4 & \mbox{, if $\Lambda < T$} \label{ma}
\end{array} \right. \mbox{ ,}
\ee
where $m_a^0 = 6.2 \times 10^{-3} \mbox{~GeV}/f_a$,  $\Lambda = 200$~MeV and $T$ always refers to the thermal bath temperature.

In order to solve the strong CP problem, the axion field must have effective couplings to the $SU(3)$ gauge fields
of the form shown in Eq.~(\ref{L_ax}). Although other non-minimal interactions are possible, they are strongly model dependent
and will be neglected here.

The supersymmetric version of Eq.~(\ref{L_ax}) implies the following couplings for the saxion and axino fields:
\be
{\cal L}_{eff} = \frac{\alpha_s}{8\pi}\frac{s(x)}{f_a} (G_{\mu\nu}^a G^{a\mu\nu} + 2i\bar{\tg}\gamma^{\mu}D_{\mu}\tg)
+  i\frac{\alpha_s}{16\pi}\frac{\bar{\tilde{a}}(x)}{f_a}\gamma_5 [\gamma^{\mu},\gamma^{\nu}] \tilde{g} G_{\mu\nu} +{\cal O}(\alpha_s^{3/2}) . 
\label{L_eff}
\ee
In most models, the axino also couples to the $U(1)_Y$ gauge boson and gaugino:
\be
{\cal L}_{eff}^{\ta} = i\frac{\alpha_Y C_{aYY}}{16\pi}\frac{\bar{\tilde{a}}(x)}{f_a}\gamma_5 [\gamma^{\mu},\gamma^{\nu}] \tilde{B} B_{\mu\nu}\,, \label{L_eff2}
\ee
where $C_{aYY}$ is a model dependent constant of order 1. 
Couplings between the saxion and axion, as well as between axinos and fermions-sfermions can also exist, but are model dependent.

If supersymmetry is unbroken, both the axino and saxion are degenerate with the axion field, hence massless, except for the tiny QCD anomaly contribution. 
However, once SUSY is broken the saxion field will receive a {\it soft} mass of order $m_{3/2}$.
On the other hand, being the fermion component of a chiral superfield,
the axino remains massless at tree level. Nonetheless, the axino can receive loop corrections to its mass\cite{maxino}
of order $m_{3/2}^2/f_a \sim 1$ eV, for $f_a \sim M_{GUT}$. In this case the axino is the lightest supersymmetric
particle (LSP) in the PQMSSM model.  However, depending on the PQMSSM model, $m_{\ta}$ can
also receive ${\cal O}(m_{3/2})$ contributions from the supergravity potential\cite{cl}. 
In this case, the axino might remain as the LSP, or the lightest neutralino could be the LSP\cite{ckls,blrs}, 
with $m_{\tz_1}<m_{\ta}$. The case of a gravitino as LSP is considered in Ref. \cite{hall}.

In the case of axino LSP, the $\tz_1$ can be long-lived due to its $1/f_a$ suppressed couplings to the axino;
this has important cosmological implications.
 Assuming the minimal interactions of Eq's.~(\ref{L_eff}) and (\ref{L_eff2}), we have the following decay rates:
\begin{eqnarray}
\Gamma(\tz_1\to\ta +\gamma) & = & \frac{\alpha^2_Y C^2_{aYY}\cos^2\theta_w v_4^{(1)2}}{128\pi^3(f_a)^2}
m_{\tz_1}^3 (1-\frac{m_{\ta}^2}{m_{\tz_1}^2})^3 \\
\Gamma(\tz_1 \to\ta +Z) & = & \frac{\alpha^2_Y C^2_{aYY}\sin^2\theta_w v_4^{(1)2}}{128\pi^3(f_a)^2}
m_{\tz_1}^3\lambda^{1/2}(1,\frac{m_{\ta}^2}{m_{\tz_1}^2},\frac{m^2_Z}{m_{\tz_1}^2}) \nonumber \\
& &\cdot\left\{\left(1-\frac{m_{\ta}^2}{m_{\tz_1}^2}\right)^2+3\frac{m_{\tz} m^2_Z}{m_{\tz_1}^3}-
\frac{m^2_Z}{2m_{\tz_1}^2}\left(1+\frac{m_{\tz}^2}{m_{\tz_1}^2}+\frac{m^2_Z}{m_{\tz_1}^2}\right)\right\}  \nonumber  \label{decays}
\end{eqnarray}
where $v_4^{(1)} = \langle \tz_1|\tB \rangle$ is the bino component of the neutralino field in the notation of Ref. \cite{wss}.
If the neutralino is the LSP, the axino will be long-lived instead. The decay rates for $\ta \to \tz_1 + \gamma/Z$ are given
by the above equations with $m_{\tz_1} \leftrightarrow m_{\ta}$. 
However, if the axino is heavier than neutralinos or gluinos, new decay modes 
$\ta\to\tz_i\gamma,\ \tz_i Z$ or $\tg g$ open up. Since the decay rates for these are discussed in Ref.\cite{blrs}, we do not reproduce them
here.

Due to the first interaction term in Eq.~(\ref{L_eff}), the saxion decay width to $gg$ is model independent, 
with decay width given by:
\be
\Gamma(s \to gg) = \frac{\alpha_s^2m_s^3}{32\pi^3 (f_a)^2} . 
\label{sdecays}
\ee
Saxions also decay to gluino pairs, but its width is always well below that to gluon pairs, as shown in Ref. \cite{ay}.
The saxion might  also decay directly into two axions. The decay width to axion pairs is given by:
\be
\Gamma(s\to aa) = \frac{k^2}{32 \pi f_a^2}m_s^3
\ee
where $k$ is a model dependent coupling. For our subsequent analysis, 
we will assume that the above decay mode is suppressed with respect to the one in Eq.~(\ref{sdecays});
this suppression is common in models with universal soft SUSY breaking terms\cite{kns}. 

\subsection{PQMSSM cosmology}

The cosmology of PQMSSM models is very rich and has to be carefully examined. 
Here, we will assume that the PQ symmetry breaks before the end of inflation so as to avoid domain wall problems\cite{dwall}.

Due to their suppressed interactions,
the axion, axino and saxion rapidly decouple from the thermal bath in the early universe or are already produced out of equilibrium,
if the reheat temperature after inflation is smaller than the decoupling temperature\cite{rtw}:
\be
T_{dcp} = 10^{11}\mbox{~GeV}\left(\frac{f_a}{10^{12} \mbox {~GeV}}\right)^2 \left(\frac{0.1}{\alpha_s}\right)^3 .
\ee
However, for the values of $f_a$ considered here ($f_a > 10^{13}$~GeV), the decoupling temperature is always
above $f_a$. Since we only consider the case where the PQ symmetry is broken before inflation ends ($T_R < f_a$), 
we always have $T_R < T_{dcp}$ and axions, saxions and axinos are never in thermal equilibrium.
In this case, the axion, saxion and axino thermal yields at $T\ll T_R$ are estimated as\cite{thermal,saxprod,kns,strumia}:
\begin{eqnarray}
Y_{a}^{TP} & \simeq &  18.6 g_s^6 \ln\left(\frac{1.501}{g_s}\right)\left(\frac{T_R}{10^{14}\; {\rm~GeV}}\right)\left(\frac{10^{12}\; {\rm~GeV}}{f_a}\right)^2 \nonumber\\
Y_{s}^{TP} & \simeq & \left(\frac{T_R}{10^{14}\ {\rm~GeV}}\right)\left(\frac{10^{12}\ {\rm~GeV}}{f_a}\right)^2  \label{yields} \\
Y_{\ta}^{TP} & \simeq & 9.2 g_s^6 \ln\left(\frac{3}{g_s}\right) \left(\frac{T_R}{10^{14}\ {\rm~GeV}}\right) \left(\frac{10^{12}\ {\rm~GeV}}{f_a}\right)^2 \nonumber
\end{eqnarray}
where $g_s$ is the strong coupling constant at $T=T_R$  and we have used $g_*(T_{dcp}) = g_*(T_R) = 229$.

In an analogous way, gravitinos are thermally produced in the early universe with yield given by\cite{gravprod}:
\be
Y_{\tG}^{TP} = \sum_{i=1}^{3}y_i g_i^2(T_R)
\left(1+\frac{M_i^2(T_R)}{3m_{\tG}^2}\right)\ln\left(\frac{k_i}{g_i(T_R)}\right)
\left(\frac{T_R}{10^{10}\ {\rm GeV}}
\right) \label{yieldG},
\ee
where $y_i=(0.653,1.604,4.276)\times 10^{-12}$, $k_i=(1.266,1.312,1.271)$, $g_i$ are the 
gauge couplings evaluated at $Q=T_R$ and $M_i$ are the gaugino masses also evaluated 
at $Q=T_R$. To compute the gravitino yield we assume $m_{\tG} \gtrsim M_i$, since we only
consider cases with $m_{\tG} \gtrsim 1$~TeV. However, most of our results are weakly dependent on this assumption.
If entropy is always conserved from $T_R$ to $T_0$, Eq's. (\ref{yields}) and (\ref{yieldG})
are still valid at $T=T_0$ and can be used to compute the relic energy density today.

Besides being produced from scattering of particles in the thermal bath, the saxion and axion fields can also contribute
to the energy density through coherent oscillations. For $T\gg m_s (m_a)$, the expansion rate of the universe
is too large and suppresses any oscillations, which only start at $3H \sim m_s (m_a)$.
The axion and saxion oscillation temperatures are then defined by\cite{vacmis,Turnerosc}:
\be
3H(T_a) = m_a(T_a)\;\; {\rm and } \;\; 3H(T_s) = m_s \label{oscond}
\ee
where $m_a(T)$ is the temperature dependent axion mass.
In most regions of the PQMSSM parameter space, we have $T_a \sim 1$~GeV and $T_s \sim 10^{10}$~GeV for a TeV scale 
saxion.\footnote{For saxion oscillation in a radiation dominated universe, $T_s=\sqrt{\frac{10^{1/2}m_sM_P}{\pi g_*^{1/2}(T_s)}}$.}
For reheat temperatures $T_R$ smaller than $T_s$, the saxion field starts to oscillate during the 
inflaton-dominated universe and will be diluted due to the inflaton entropy injection. 
In this case, the saxion coherent oscillation density increases
with $T_R$ until $T_R > T_s$, where it then becomes $T_R$ independent.
The specific details of the transition from the static to the
oscillating regime are strongly model dependent, but can be parametrized by an arbitrary initial field amplitude.
If the axion oscillation starts in a radiation dominated universe, for $T\ll T_{a,s}$,
the axion and saxion coherent oscillation Yields are given by\footnote{Our saxion Yield expressions
differ from Ref.\cite{kns} by a numerical factor because they assume the saxion field begins to oscillate at $H=m_s$
while we take $3H=m_s$ to be consistent with the usual axion oscillation condition.}:
\begin{eqnarray}
Y_{a}^{RD} & \simeq & \left\{ \begin{array}{ll}
4.8\times 10^{-9}f(\theta_i)\theta_i^2 g_*^{-5/12}(T_a)\left(\frac{f_a}{10^{12}\ {\rm~GeV}}\right)^{7/6} & \mbox{, if $\Lambda < T_a$} \\
3.4\times 10^{-11} f(\theta_i)\theta_i^2 g_*^{-1/4}(T_a)\left(\frac{f_a}{10^{12}\ {\rm~GeV}}\right)^{3/2} & \mbox{, if $T_a<\Lambda$}
\end{array} \right.\\
Y_{s} & \simeq & \left\{ \begin{array}{ll}
7.8\times 10^{-5}\ {\rm~GeV}\left(\frac{1\ {\rm~GeV}}{m_s}\right)^{1/2}
\left(\frac{f_a}{10^{12}\ {\rm~GeV}}\right)^2\left(\frac{s_i}{f_a}\right)^2 & \mbox{, if $T_s<T_R$} \\
 1.9\times 10^{-8}\ {\rm~GeV}\left(\frac{T_R}{10^5m_s}\right)
\left(\frac{f_a}{10^{12}\ {\rm~GeV}}\right)^2\left(\frac{s_i}{f_a}\right)^2 & \mbox{, if $T_R < T_s$}
\end{array} \right. \label{yieldco}
\end{eqnarray}
where\cite{vg1} $f(\theta_i)=\left[\ln\left(\frac{e}{1-\theta_i^2/\pi^2}\right)\right]^{7/6}$ and
$s_i$ and $\theta_i f_a$ are the initial saxion and axion field amplitudes and are expected to be of order $f_a$.

From the above expressions we see that for most purposes the PQMSSM parameter space can be restricted to:
\be
\left\{f_a,\; m_{\ta},\; m_s,\; s_i,\; \theta_i,\; T_R \right\} \mbox{ + SUSY spectrum} .
\ee
For simplicity, we take $C_{aYY}=8/3$ as in the DFSZ\cite{dfsz} axion model or in the KSVZ\cite{ksvz} 
model with PQ quark charges $2/3$.

\subsubsection{Early saxion dominated universe}
\label{sec:SD}

Due to the small couplings of the axion supermultiplet, its components decouple
from the thermal bath at very high temperatures.
For temperatures below $m_s$ for thermally produced saxions or below $T_s$ for
coherent oscillating saxions, the energy density of the saxion field scales as $R^{-3}$.
Since the radiation energy density scales as $R^{-4}$, if
the saxion $s$ is sufficiently long-lived, it will dominate the energy
density of the universe. Defining $T_e$ and $T_D$ as the temperatures at which
the saxion dominated era starts and ends, respectively, we have:
\begin{eqnarray}
\rho_s(T_e) = \rho_{\gamma}(T_e) \To T_e & = & \frac{4}{3} m_s Y_s \label{TE} \\
\Gamma_s = H(T_D) \To T_D & = & \frac{\sqrt{\Gamma_s M_P}}{(\pi^2g_*(T_D)/90)^{1/4}} \label{TD}
\end{eqnarray}
where $\Gamma_s$ is the saxion decay rate given by Eq.~(\ref{decays}) and $M_P$ is the reduced Planck mass.
If $T_D > T_e$, the saxion decays before dominating the energy density and
its effects can be safely neglected\footnote{If the saxion field has a large branching ratio into axions,
its decays contribute to a hot DM component even for $T_e < T_D$. However, as mentioned in Sec.~\ref{sec:pheno},
here we assume $BR(s\to gg) \simeq 1$.}.
However, if $T_D < T_e$,
the universe becomes matter (saxion) dominated for $T_D < T < T_e$. At $T = T_D$
most of the saxions have decayed and the radiation dominated era
resumes. 

Fig.~\ref{fig:Teplot} shows $T_e$ as a function of the reheat temperature
for $m_{s} = 20$~TeV, $s_i = f_a$ and $f_a = 10^{15}$, $10^{16}$ and $10^{17}$~GeV.
As we can see, $T_e$ increases with $T_R$ until $T_s < T_R$, when the saxion starts to
oscillate after inflation. In this regime, the saxion Yield and hence $T_e$ becomes
independent of $T_R$. From Fig.~\ref{fig:Teplot}, we see that for $f_a \sim 10^{16}$~GeV
and $m_{s} = 20$~TeV, $T_e$ can be as large as $10^6$~GeV.

%
\FIGURE[t]{
\includegraphics[width=10cm]{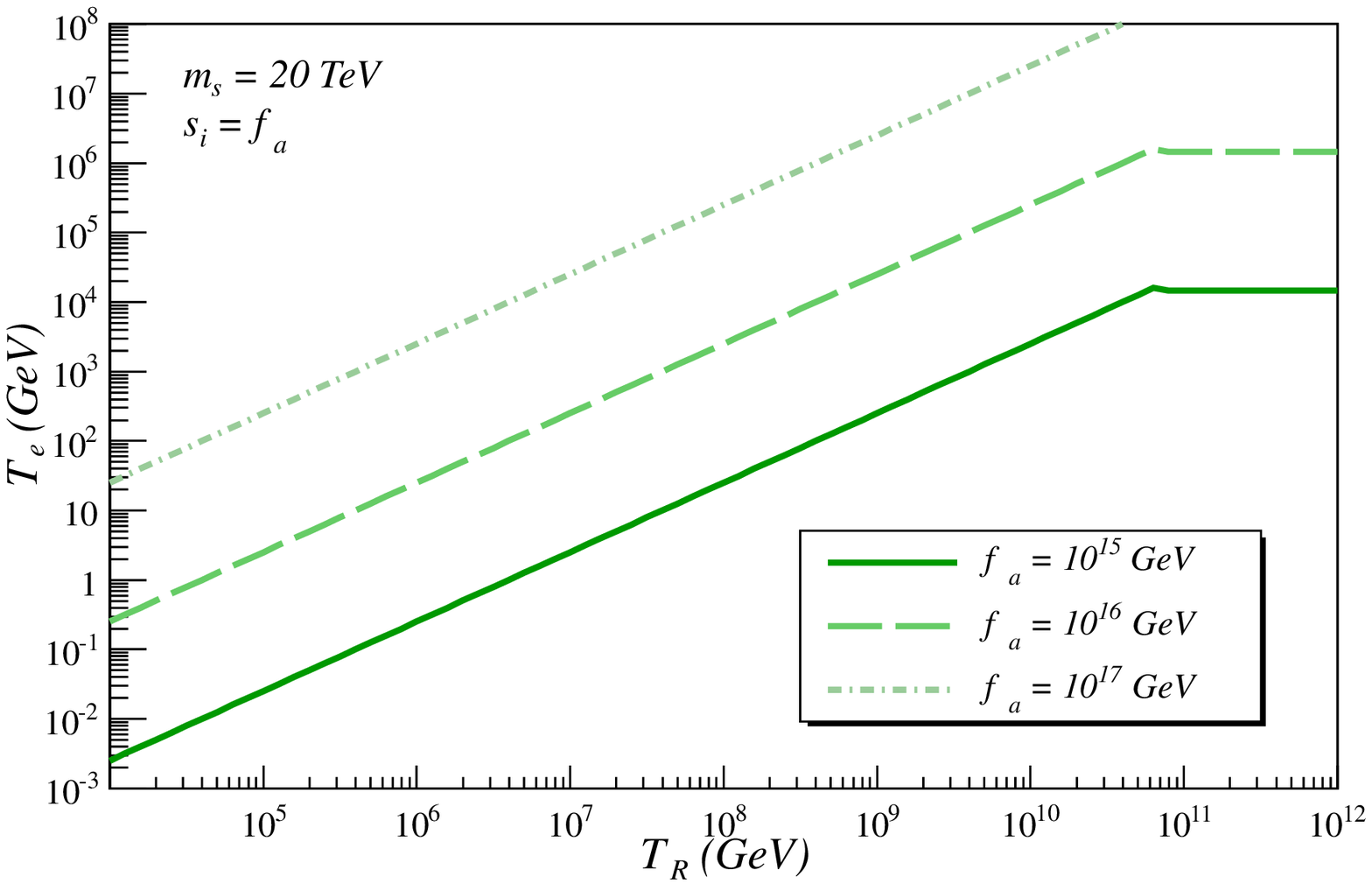}
\caption{Saxion-radiation equality temperature ($T_e$) versus the reheat
temperature after inflation ($T_{R}$)
for $f_a = 10^{15}$, $10^{16}$ and $10^{17}$~GeV (bottom to top), $m_{s} = 20$~TeV and $s_i = f_a$.}
\label{fig:Teplot}}

In Fig.~\ref{fig:Tsplot}, we plot $T_D$ as a function of $m_{s}$ for $f_a = 10^{15}$, $10^{16}$ and $10^{17}$~GeV.
We see that $T_D$ can span a wide range of values, with $T_D \lesssim 0.1$~MeV for a sub-TeV saxion and $f_{a}\agt 10^{15}$~GeV.
However, if $T_D < 5$~MeV, entropy will be injected during the neutron freeze-out
and the neutron-proton ratio will be significantly diluted. Since the success of BBN predictions strongly constrains
the neutron-proton ratio to the value obtained in a radiation dominated universe, we must
have $T_D > 5$~MeV in order to have a neutron freeze-out during the radiation dominated era, thus preserving the
successful BBN predictions. 
As shown in Fig.~\ref{fig:Tsplot}, this requires $m_s \gtrsim 50$~TeV for $f_a \sim 10^{16}$~GeV.
This bound can be softened by including additional model-dependent saxion decays such as $s\to aa$.

%
\FIGURE[t]{
\includegraphics[width=10cm]{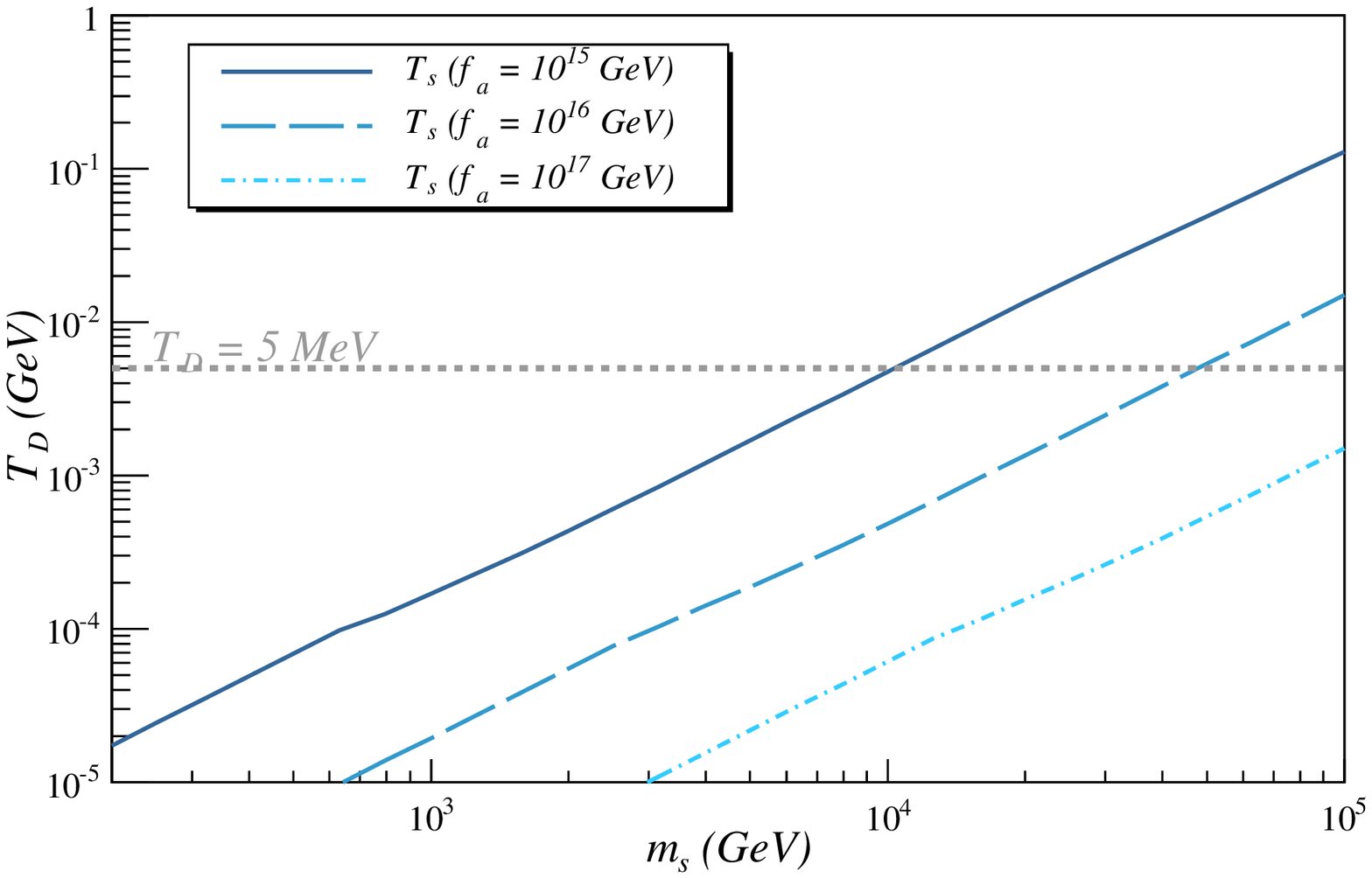}
\caption{Saxion decay temperature ($T_D$) versus the saxion mass ($m_{s}$)
for $f_a = 10^{15}$, $10^{16}$ and $10^{17}$~GeV (top to bottom).
We also show in grey the BBN constraint on $T_D$ ($T_D > 5$~MeV).}
\label{fig:Tsplot}}

Once $s$ starts to decay, entropy will be injected and effectively
dilute all other relic densities already decoupled from the thermal bath,
such as axinos, axions, gravitinos and possibly neutralinos. If the decoupling
happens {\it before} the saxion dominated era (this is always the case for thermal axions, axinos and gravitinos),
the dilution factor ($r$) is approximately given by\cite{turner}:
\be
r=\frac{S_f}{S_i}\simeq 0.82\bar{g}_*^{1/4}\frac{Y_s m_s}{(M_{P}\Gamma_s)^{1/2}} \approx \frac{T_e}{T_D}, \label{r}
\ee
where $\bar{g}_*$ is the number of relativistic degrees of freedom averaged over the saxion decay
period, which we approximate by $g_{*}(T_D)$. Fig.~\ref{fig:Rplot} shows contours of $r$ values in the $T_R$ vs $f_a$ plane.
We see that, for $s_i = f_a$ and $f_a \sim 10^{16}$~GeV, values of $r$ larger than $10^4$ are easily obtained.

%
\FIGURE[t]{
\includegraphics[width=10cm]{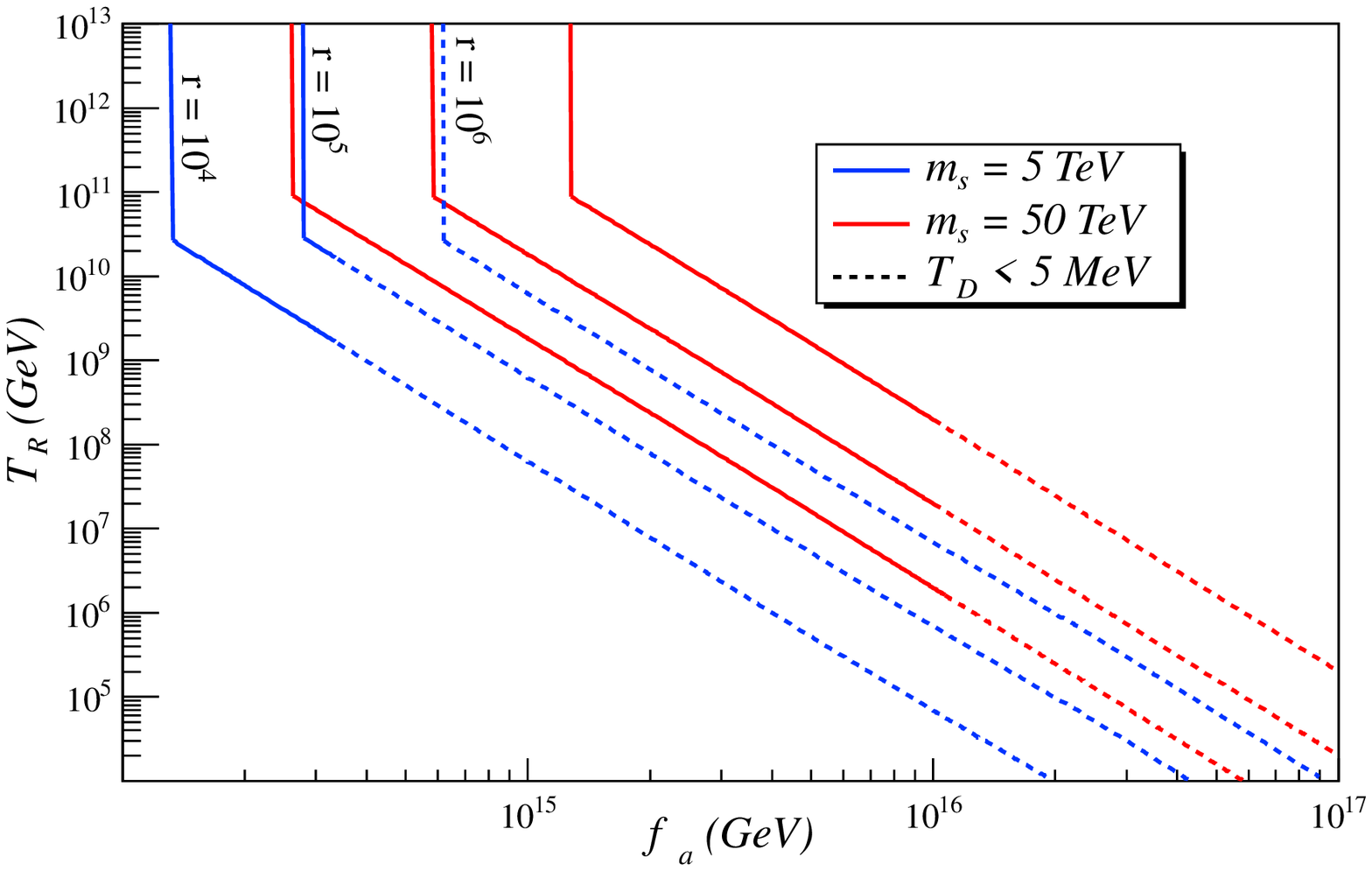}
\caption{Entropy dilution factor ($r$) contours in the $f_a$-$T_R$ plane for
$m_{s} = 5$ and 50 TeV. The curves have $r$ values $10^4$, $10^5$ and $10^6$ from bottom to top.
The dashed region is excluded by the BBN constraints on $T_D$.}
\label{fig:Rplot}}

However, if the decoupling from the thermal bath happens {\it during} the saxion dominated period,
as it might be the case for neutralinos or for the beginning of axion oscillation, the dilution
factor will be modified due to the faster expansion rate during the saxion dominated
phase. As a result, the neutralino freeze-out temperature ($T_{fr}$) and the axion oscillation
temperature ($T_a$) will be modified, if $T_D < T_{a,fr} < T_e$. 

Finally, if the decoupling (or axion oscillation) happens {\it after} the saxion decay,
there will be no dilution. In summary, for the case of an early saxion dominated universe ($T_e > T_D$),
we have:
\be
\Omega_x = \left\{ \begin{array}{ll}
\Omega_x^{RD}/r \;  \mbox{, if $T_e < T_x$} \\
\Omega_x^{MD}\;  \mbox{, if $T_{DD} < T_x < T_e$} \\
\Omega_x^{DD}\;  \mbox{, if $T_D < T_x < T_{DD}$} \\
\Omega_x^{RD}\;  \mbox{, if $T_x < T_D$}
\end{array} \right. \label{omegas}
\ee
where $\Omega_x$ represents the axion, axino, gravitino or neutralino relic density, $\Omega_x^{RD}$ is
the corresponding relic density in a radiation dominated (RD) universe (such as when $T_D > T_e$),
$T_x$ is the decoupling or axion oscillation temperature and $T_{DD}$ marks the transition
between the matter dominated phase (MD) and the decaying particle dominated phase (DD) and
can be approximated by (see Appendix):
\be
T_{DD} = \left(\frac{g_*(T_D)}{g_*(T_x)} T_e T_D^4\right)^{1/5} .
\ee

%
\FIGURE[t]{
\includegraphics[width=12cm]{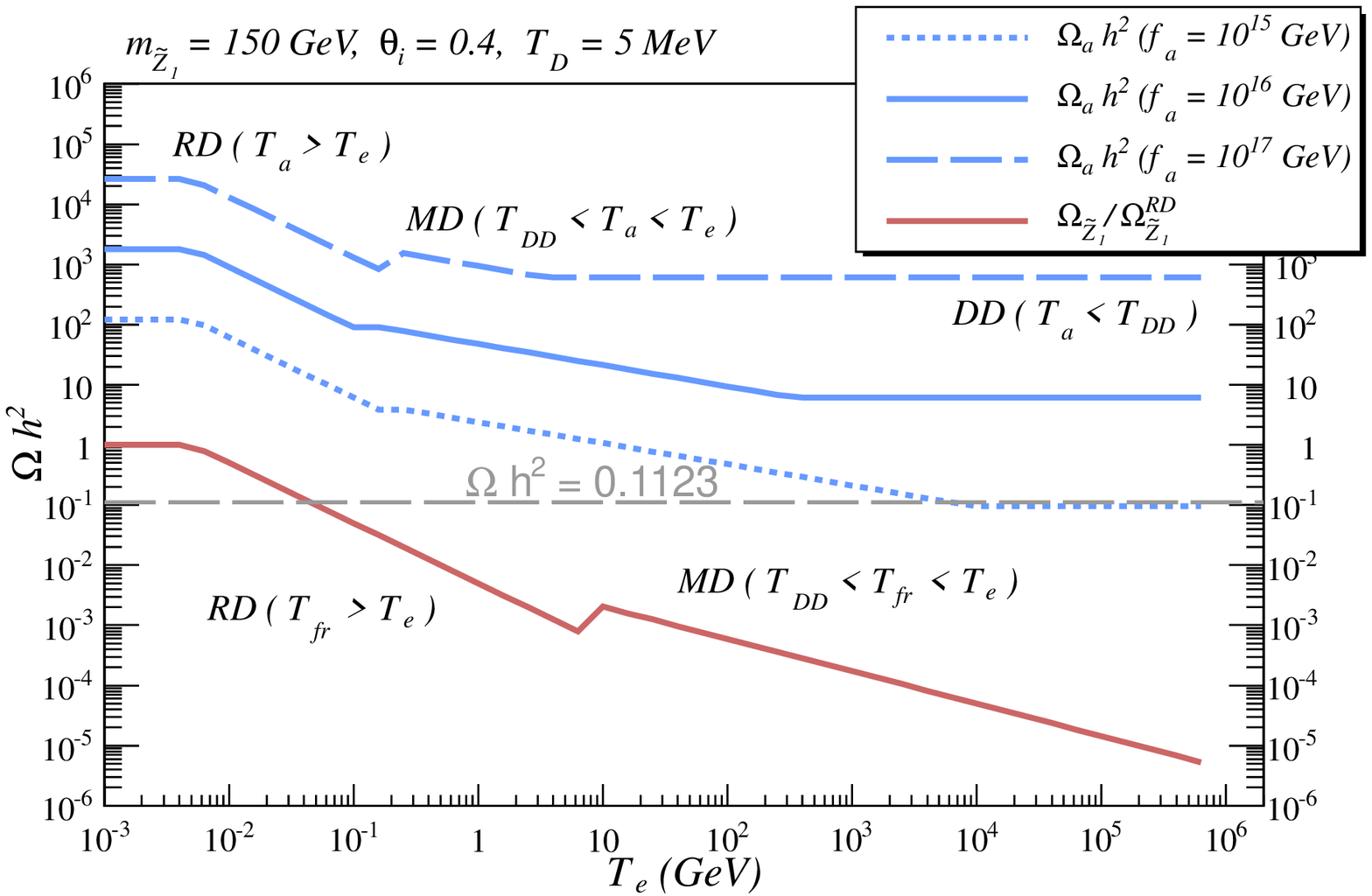}
\caption{The axion and neutralino relic density versus $T_e$, the temperature
at which the universe becomes saxion dominated. The saxion decay temperature is fixed at its
minimum value allowed by BBN (5 MeV), $m_{\tz_1}$ = 150~GeV and $\theta_i = 0.4$. The
dotted, solid and dashed blue lines corresponds to the axion relic density for 
$f_a = 10^{15}$, $10^{16}$ and $10^{17}$~GeV, respectively. We also show the regions
where the axion (neutralino) starts to oscillate (decouple) during the radiation (RD),
matter (MD) or decaying particle (DD) dominated era. The neutralino relic density
is normalized by its MSSM value.}
\label{fig:omegas}}

As mentioned before, the axino and gravitino fields will always have $T_x \gg T_e$, so
their relic densities will simply be diluted by $r$. On the other hand the neutralino freeze-out
temperature and the axion oscillation temperature may fall in any of the intervals in Eq.~(\ref{omegas}).
The appropriate expressions for $T_{a}$, $T_{fr}$, $\Omega_{a}$ and $\Omega_{\tz_1}$ for the RD, MD and DD
cases are given in the Appendix.
To illustrate the behavior of $\Omega_{\tz_1}$ and $\Omega_a$ in the distinct regions of Eq.~(\ref{omegas}),
we show in Fig.~\ref{fig:omegas} the relic densities versus $T_e$, with
$T_D = 5$~MeV, $\theta_i = 0.4$ and $m_{\tz_1} = 150$~GeV, for $f_a = 10^{15}$, $10^{16}$ and $10^{17}$~GeV.
 As we can see, for small values of $T_e$ ($< T_{a,fr}$), the
axion and neutralino relic densities are simply diluted by $1/r$. Once $T_e \sim 0.1$ (10)~GeV, the
axion (neutralino) starts to oscillate (decouple) during the matter dominated (MD) era. As a result,
the relic density is no longer diluted by $1/r$, but by a smaller factor. Once $T_e \gg T_a$, the axion
starts to oscillate during the decaying particle dominated phase (DD) and becomes independent of $T_e$,
despite the increase of entropy injection. Due to its
large freeze-out temperature ($T_{fr} \sim 7$~GeV),
the neutralino never decouples during the DD phase for the range of $T_e$ values shown.

\section{The $f_a = M_{GUT}$ regime}
\label{sec:bigfa}

Assuming $f_a$ to be of order the GUT scale ($\sim 10^{16}$~GeV) has several
important consequences for the PQMSSM cosmology:
\bi
\item The thermal production of all the components of the axion supermultiplet will
be strongly suppressed  by the large value of $f_a$ (see Eq.~(\ref{yields})).\footnote{
The thermal production could still be relevant for $T_R > f_a$. 
However, for such large $T_R$ values the PQ symmetry would
 only be broken after inflation, resulting in a universe broken into
domains with different values of $\theta_i$ and $s_i$. Since this scenario
requires a different dark matter treatment, we assume $T_R < f_a$.}
Therefore the axino and thermal axion contributions to the DM density are negligible and the
saxion and axion densities are dominated by the coherent oscillation component.
\item Assuming $s_i \sim f_a$, large $f_a$ will result in a large energy density for the coherent
oscillating saxion field as shown in Eq.~(\ref{yieldco}).
\item Since $T_e \propto f_a^2$ (for coherent oscillating saxions) and $T_D \propto 1/f_a$,
large $f_a$ will usually result in a large entropy injection $r=\frac{T_e}{T_D}$ from saxion decays, as shown in 
Figs.~\ref{fig:Rplot} and \ref{fig:omegas}.
\item The axion field will be extremely light ($\sim 10^{-10}$ eV).
\item Since $\Gamma_{s} \propto 1/f_a^2$, saxions will be long-lived,
potentially spoiling the BBN predictions.
\item For an axino (neutralino) LSP, $\Gamma_{\tz_1 (\ta)} \propto 1/f_a^2$ and the
neutralino (axino) will be long-lived and a threat to successful BBN.
\ei

From the above points, we see that $f_a \sim M_{GUT}$ naturally leads to 
a long-lived saxion with large energy density from coherent saxion oscillations. As shown
in Sec.~\ref{sec:SD}, this results in an early saxion dominated
universe with a large dilution of other relics, due to the entropy injection during saxion decays.
Below we discuss the implications of this scenario for big-bang nucleosythesis and the dark
matter relic density.

\subsection{The dark matter constraint}
\label{sec:DMconst}

If the saxion field is neglected, the axion relic density is given by
\be
\Omega_a h^2 = \left\{ \begin{array}{ll}
9.23\times10^{-3} \theta_i^2 f(\theta_i)  \frac{1}{g_*(T_a)^{1/4}} \left(\frac{f_a}{10^{12}\ {\rm GeV}}\right)^{3/2} \mbox{, if $T_a < \Lambda$} \\
1.32\ \theta_i^2 f(\theta_i) \frac{1}{g_*(T_a)^{5/12}}\left(\frac{f_a}{10^{12}\ {\rm GeV}}\right)^{7/6} \mbox{, if $T_a > \Lambda$}
\end{array} \right.
\ee
where $\Lambda = 200$~MeV. Since the mis-alignment angle $\theta_i$ is supposed to be a random variable in the interval $[0,\pi]$,
it is usually assumed that $\theta_i \sim 1$. Thus, the constraint on the dark matter relic density
\be
\Omega_{DM} h^2 = 0.1123 \pm 0.0035 \; \mbox{( at 68\% CL)}
\ee
implies $f_a \lesssim 10^{12}$~GeV. In this context the PQ and GUT scales can only be unified if
$\theta_i$ takes unnaturally small values ($\lesssim 10^{-3}$).

However, as shown in the Appendix,
once the entropy injection from saxion decays is included, we have (for $T_a \gtrsim \Lambda$):
\be
\Omega_a \propto \left\{ \begin{array}{ll}
 f_a^{7/6} \times T_D/T_e\;  \mbox{, if $T_e < T_a$} \\
 f_a^{14/11} \times T_D/T_e^{4/11}\;  \mbox{, if $T_{DD} < T_a < T_e$} \\
 f_a^{3/2} \times T_D^2\;  \mbox{, if $T_D < T_a < T_{DD}$} .
\end{array} \right. \label{omegas2}
\ee
Since $T_D \propto 1/f_a$ and $T_e \propto f_a^2$ (for coherent oscillating saxions), from Eq.~(\ref{omegas2})
we see that the axion relic density actually {\it decreases} with $f_a$. In this case, the $f_a \lesssim 10^{12}$~GeV bound
can be potentially avoided. On the other hand, if $T_a \lesssim \Lambda$, $\Omega_a$ {\it increases}
with $f_a$, unless $T_a > T_e$. In both cases we see that $\Omega_a h^2$ is maximally suppressed for $T_D = T_D^{min} = 5$~MeV.
From Fig.~\ref{fig:omegas}, we see that the maximum dilution occurs in the DD regime with $T_a \gtrsim \Lambda$.
Using the expression for $\Omega_a^{DD} h^2$
in the Appendix, we estimate the maximum $\theta_i$ value allowed by the DM constraint as:
\be
\Omega_a^{DD} h^2 < 0.11 \To \theta_i < \theta_i^{max} \simeq (69-106) \left(\frac{5\ {\rm MeV}}{T_D}\right)
\left(\frac{10^{12}\ {\rm GeV}}{f_a}\right)^{3/4} . 
\label{thetamax}
\ee
The uncertainty on $\theta_i^{max}$ comes from the uncertainty on the axion mass at $T_a \sim \Lambda$ 
(see Eq.~(\ref{ma})).

\subsection{BBN bounds}
\label{sec:bbn}
Since we will first assume a PQMSSM with a neutralino NLSP and an axino LSP, the neutralino will
decay into axinos and SM particles and can be long-lived for $f_a\sim M_{GUT}$.
Furthermore, the PQMSSM will also contain long-lived gravitinos and saxions.
All or any of these three fields can decay during or after BBN,
potentially spoiling its successful predictions, unless their relic densities are extremely small
at the time of their decay.

The $\tz_1$ decay width and hadronic branching fraction are calculated in Ref.~\cite{tr}.
Fig.~\ref{fig:Z1lt} shows the neutralino lifetime for a bino-like $\tz_1$,
as a function of $m_{\tz_1}$, for $f_a = 10^{15}$, $10^{16}$ and $10^{17}$~GeV  and $m_{\ta} \ll m_{\tz_1}$.
Usual BBN bounds on late decaying particles require\cite{jedamzik}
\be
 \tau \lesssim 0.01s\ \mbox{  if $\Omega h^2 \gg 1$ or }\; \Omega h^2 \lesssim 10^{-4}\ \mbox{  if $\tau \gtrsim 10^4s$} .
\ee
From Fig.~\ref{fig:Z1lt}, we see that unless $m_{\tz_1}$ is in the multi-TeV range,
the neutralino life-time ($\tau_{\tz_1}$) will be well above $10^4$s. To maintain sub-TeV values of $m_{\tz_1}$,
extremely small values of $ \Omega_{\tz_1} h^2$ are required in order to satisfy the BBN
constraints. Since in almost all of the MSSM parameter space 
$10^{-3} < \Omega_{\tz_1}^{std} h^2 < 10^{3}$\cite{sug19},
the BBN constraints would require an enormous fine-tuning of the MSSM parameters.
However, if neutralinos decouple from the thermal bath before saxions have decayed, their
relic density will also be diluted by the saxion decay, according to Eq.~(\ref{omegas}).
As seen in Fig.~\ref{fig:omegas}, the neutralino dilution can
exceed $10^5$ for large enough $T_e$.  Hence,
the BBN bounds on late $\tz_1$ decays can be potentially
avoided due to the large suppression of $\Omega_{\tz_1} h^2$, without
the need for fine-tuned MSSM parameters.

%
\FIGURE[t]{
\includegraphics[width=10cm]{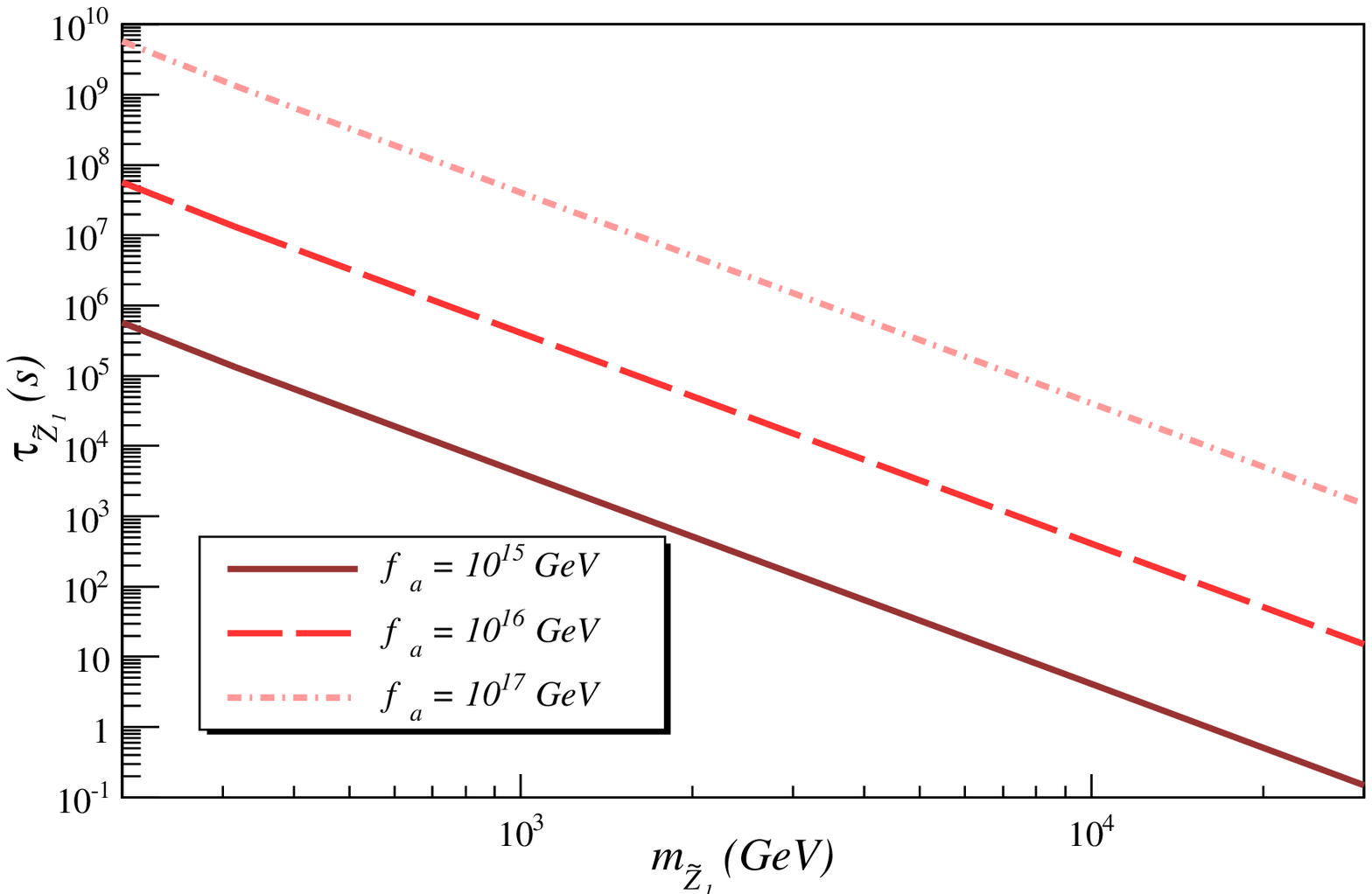}
\caption{The neutralino lifetime as a function of the neutralino mass
for $f_a = 10^{15}$, $10^{16}$ and $10^{17}$~GeV (bottom to top) and
$m_{\ta} \ll m_{\tz_1}$.}
\label{fig:Z1lt}}

Thermal gravitinos are produced out of equilibrium via radiation off of particles
in the thermal bath (see Eq.~(\ref{yieldG})) and have decay rates suppressed by $1/M_P^2$, 
decaying during or after BBN, for $m_{\tG} \lesssim 50$~TeV\cite{moroi}.\footnote{Here, we take $m_{\tG}$ to
represent the physical gravitino mass, while $m_{3/2}$ is the Lagrangian gravitino mass parameter.}
Therefore, if $T_R$ is large enough to significantly produce gravitinos in the early universe, their
late decay will spoil the BBN predictions.
This is usually known as the {\it gravitino problem}\cite{gravprob}, common in most supergravity models
with moderate $m_{\tG}$ and reheat temperatures above $10^{5}$~GeV\cite{khlopov}. However, since
gravitinos always have decoupling temperatures larger than the reheat temperature,
the gravitino relic density is diluted by $1/r$, as shown in Eq.~(\ref{omegas}).
Since $r$ can easily exceed $10^4$ for $f_a \sim M_{GUT}$, as seen in Fig.~\ref{fig:Rplot}, the gravitino
relic density will be strongly suppressed, naturally avoiding the gravitino problem.

In the scenario where $m_{\ta} \sim m_{3/2}$ and the neutralino is the LSP,
axinos will cascade decay to neutralinos, as discussed in Ref.\cite{blrs}. In this case
the BBN bounds on late decaying axinos are easily avoided since $Y_{\ta}^{TP}$ is
suppressed for large $f_a$. Furthermore, if $m_{\ta} \gtrsim m_{\tg}$, the decay mode
$\ta\to\tg g$ considerably reduces the axino lifetime and it usually decays before BBN.

Finally, as already discussed in Sec.~\ref{sec:SD}, the BBN bounds on late decaying saxions 
require $T_D > 5$~MeV. Using Eq's~(\ref{sdecays}) and (\ref{TD}), we have:
\be
T_D > 5\ {\rm MeV} \To m_s \gtrsim 0.1\ {\rm TeV} \left(\frac{f_a}{10^{12}\ {\rm GeV}}\right)^{2/3} .
\label{msmin}
\ee
Thus, as already shown by Fig.~\ref{fig:Tsplot}, BBN bounds on saxion decays require $m_s \gtrsim 50$~TeV,
for $f_a \sim 10^{16}$ GeV.

\section{Results for an axino LSP}
\label{sec:axino}

In this Section, we discuss under which
values of PQMSSM parameters we can conciliate $f_a \sim M_{GUT}$ with the dark matter and BBN constraints for
the case of a light axino as LSP and present explicit examples for our previous discussion.
In this scenario, for large $f_a$, the total dark matter relic density will be given by\cite{bs}:
\be
\Omega_{DM} h^2 = \Omega_a h^2 + \Omega_{\ta}^{TP} h^2 + \frac{m_{\ta}}{m_{\tz_1}} \Omega_{\tz_1} h^2 + \frac{m_{\ta}}{m_{\tG}} \Omega_{\tG} h^2
\ee
where we have neglected the subdominant contributions from thermal axions and the gravitino contribution to the axion relic density,
which is suppressed by $m_a/m_{\tG}$. To compute the above relic densities, we use Eqs.~(\ref{yields}), (\ref{yieldco}), (\ref{omegas})
and the expressions in the Appendix. The BBN bounds on late decaying neutralinos and gravitinos are obtained
from Figs.~9 and 10 in Ref.\cite{jedamzik}, with a linear interpolation for different values of $m_{\tz_1}$ or $m_{\tG}$.
For late decaying saxions we simply require $T_D > 5$~MeV, to avoid further entropy injection during the neutron freeze-out.

\subsection{Specific example for axino as LSP}

Fig.~\ref{fig:OmegaR} shows the axion, axino, neutralino and gravitino relic densities as a function of $f_a$.
For the PQMSSM parameters we take $m_s = 50$~TeV, $s_i = 10f_a$, $m_{\tG} = 1$~TeV, $T_R = 10^{11}$~GeV and
$m_{\ta}=0.1$ MeV.
We assume $m_{\tz_1} = 150$~GeV and the neutralino relic density {\it before} dilution to be $\Omega_{\tz_1}^{std} h^2 = 10$.
For each $f_a$ value, a different value for $\theta_i$ is chosen so that $\Omega_{DM} h^2 = 0.1123$ is satisfied.
As we can see from Fig.~\ref{fig:OmegaR}, the dilution of the axino, neutralino and gravitino relic densities rapidly
increases with $f_a$ due to the increasing rate of saxion production via oscillations. For $f_a \lesssim 7\times 10^{14}$~GeV, we have
$\Omega_{\tz_1} h^2 \gtrsim 10^{-4}$ and BBN constraints on late decaying neutralinos exclude this region. However, 
if a smaller value of $\Omega_{\tz_1}^{std}$ had been chosen, smaller $f_a$  values would be allowed.
Once $f_a \sim 10^{15}$~GeV, the entropy injection from saxion decays dilutes the neutralino relic density
to values below $10^{-4}$,
making these high $f_a$ values consistent with BBN. Finally, when $f_a \sim  10^{16}$~GeV, the saxion starts
to decay at $T=T_D < 5$~MeV and these solutions become once again excluded by the BBN constraints. We can also see
that the gravitino relic density is strongly suppressed despite the large $T_R$ value,
easily avoiding the BBN constraints on late decaying gravitinos.
Also, despite being the LSP, the axino does not significantly contribute to $\Omega_{DM} h^2$, and the
cosmologically allowed region around $f_a\sim 10^{16}$ GeV has little dependence on $m_{\ta}$.

%
\FIGURE[t]{
\includegraphics[width=10cm]{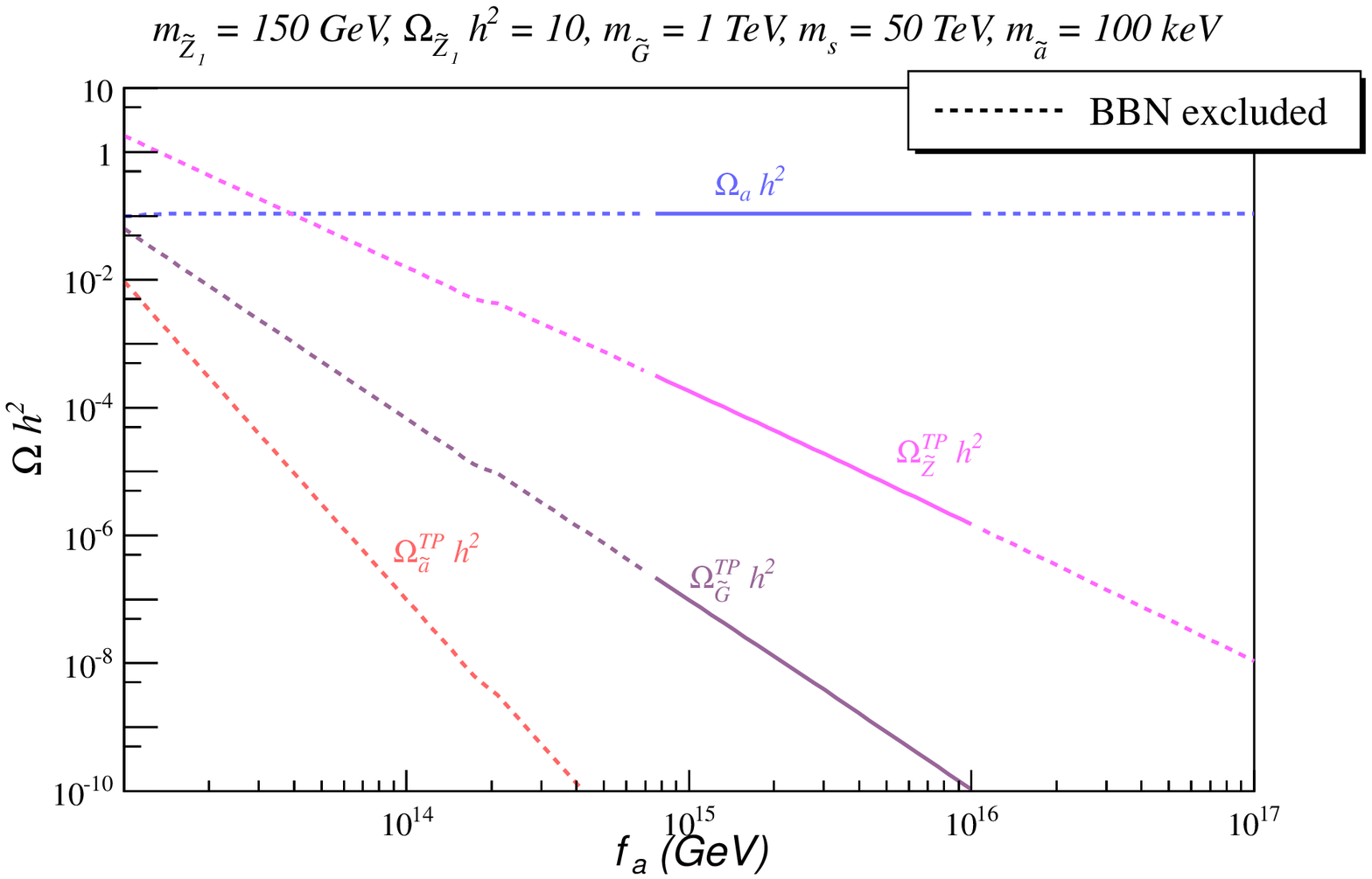}
\caption{The axion, axino, neutralino and gravitino relic densities as a function of $f_a$ for
$m_{\tz_1} = 150$~GeV, $\Omega_{\tz_1} h^2 = 10$, $m_{\ta} = 100$~keV, $m_{\tG} = 1$~TeV, $m_{s} = 50$~TeV,
$T_{R} = 10^{11}$~GeV and $s_i = 10f_a$. The misalignment angle ($\theta_i$) is chosen such as 
$\Omega_{DM} h^2 \simeq \Omega_{a} h^2 = 0.1123$.
The dashed region is excluded by BBN bounds on neutralino and gravitino late decays (for $f_a \lesssim 7\times10^{14}$~GeV)
or $T_{D} < 5$~MeV (for $f_a \gtrsim 10^{16}$~GeV).}
\label{fig:OmegaR}}

In Fig.~\ref{fig:ThetaR}, we show the values of $\theta_i$ necessary to satistfy the dark matter constraint
for the same PQMSSM parameter values used in Fig.~\ref{fig:OmegaR}. For $f_a \lesssim 10^{14}$~GeV,
the axion oscillates after the saxion has decayed ($T_a < T_D$) and $\Omega_a h^2$ is not diluted by the early
entropy injection. In this regime, the values of $\theta_i$ required to satisfy $\Omega_{a} h^2 = 0.1123$
rapidly decrease with $f_a$, since the axion relic density increases with $f_a$ for $T_a < T_D$.
For $ 10^{14}\; \mbox{ GeV} \lesssim f_a \lesssim 6\times 10^{15}$~GeV,
the axion starts to oscillate in the decaying particle dominated (DD) regime ($T_D < T_a < T_{DD}$)
 and $\Omega_a h^2$ {\it decreases} with $f_a$,
as discussed in Sec.~\ref{sec:DMconst}. As a result, $\theta_i$ increases with $f_a$, although it is still required
to be small ($\lesssim 0.07$). Once $f_a \gtrsim 6\times 10^{15}$~GeV, the axion oscillation still starts in the DD era, but now
with $T_a < \Lambda$. As shown in the Appendix, in this case $\Omega_a h^2 \propto f_a$ and the
mis-alignment angle once again decreases as $f_a$ increases, although with a smaller slope than in the RD era.

From Figs.~\ref{fig:OmegaR} and \ref{fig:ThetaR}, we see that $f_a\sim M_{GUT}$ 
can indeed be consistent with the dark matter and BBN bounds. However, for the above choice of PQMSSM parameters,
the region of parameter space consistent with all bounds is considerably
restricted. Furthermore, $\theta_i$ still has to take small values, as
would be the case in the PQ standard model cosmology, where the saxion field is neglected and
$f_a \sim M_{GUT}$ can be obtained if we take $\theta_i \lesssim 3\times 10^{-3}$\cite{vg1}.
Since the main purpose of the PQ mechanism is to avoid a huge fine-tuning in $\theta_{QCD}$,
it is desirable to avoid unnaturally small values for the mis-alignment angle as well.
With this is mind, we point out that $\theta_i$ can
take considerably larger values ($\sim 0.07$) once the saxion and axino fields are included.
Furthermore, the dilution of the neutralino and gravitino relic densities allows for an elegant
way of avoiding the BBN constraints without having to assume extremely small $\Omega_{\tz_1} h^2$,  low
reheat temperatures or a multi-TeV gravitino.

%
\FIGURE[t]{
\includegraphics[width=10cm]{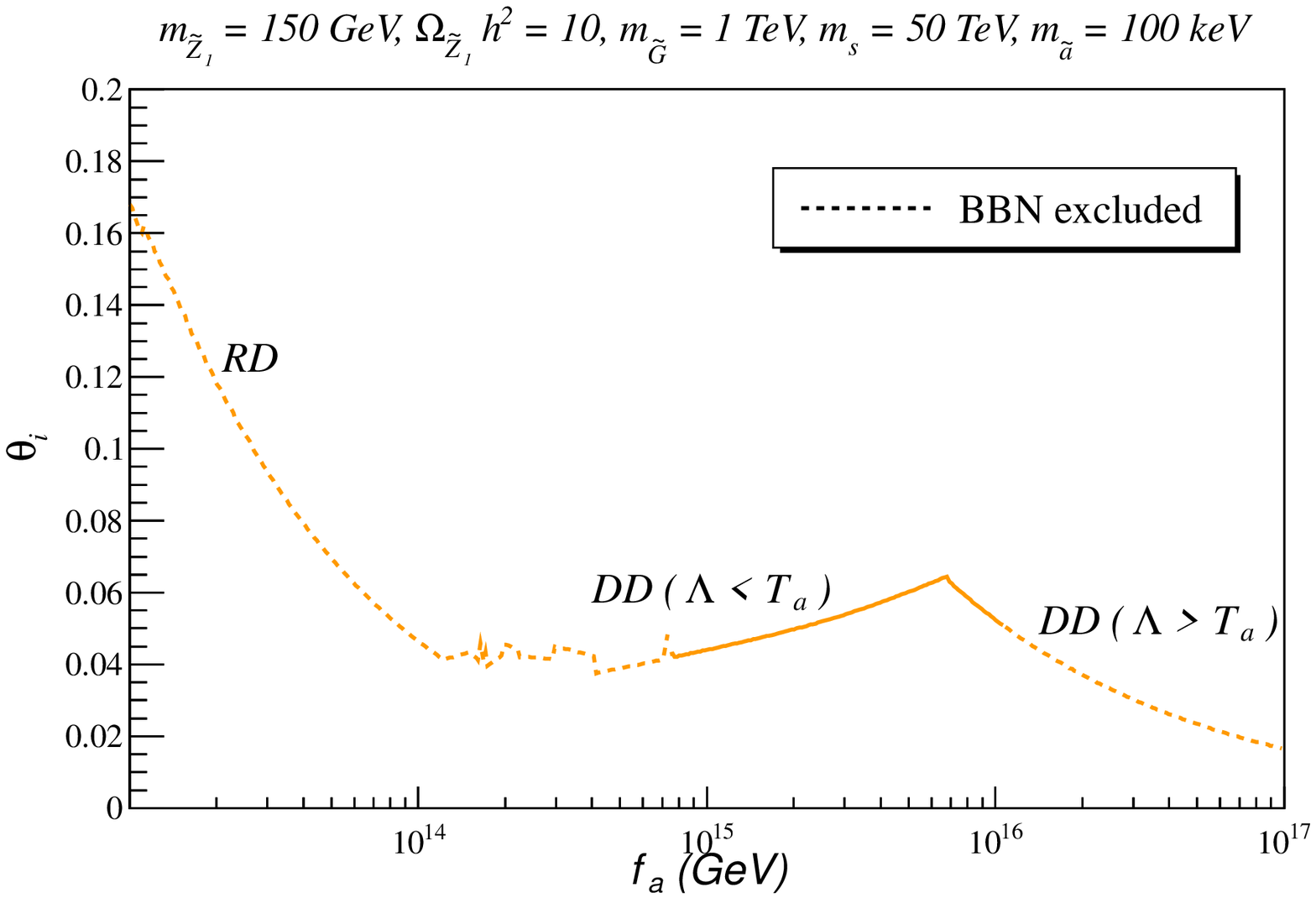}
\caption{The misalignment angle required for $\Omega_{DM}h^2 = 0.1123$ as a function of $f_a$ for
$m_{\tz_1} = 150$~GeV, $\Omega_{\tz_1} h^2 = 10$, $m_{\ta} = 100$~keV, $m_{\tG} = 1$~TeV, $m_{s} = 50$~TeV,
$T_{R} = 10^{11}$~GeV and $s_i = 10f_a$.
The dashed region is excluded by BBN bounds on neutralino and gravitino late decays (for $f_a \lesssim 7\times10^{14}$~GeV)
or $T_{D} < 5$~MeV (for $f_a \gtrsim 10^{16}$~GeV).}
\label{fig:ThetaR}}

\subsection{General results for axino LSP}

The above arguments are, however, limited by our choice of the PQMSSM
parameters used in Figs.~\ref{fig:OmegaR} and \ref{fig:ThetaR}. In order
to generalize these results,
we perform a random scan over the following parameters:
\bea
f_a    &\in& [10^{15},\;10^{17}]\ {\rm GeV}\,, \nonumber\\
m_s    &\in& [10^{3},\;10^{5}]\ {\rm GeV}\,, \nonumber\\
s_i/f_a &\in& [10^{-2},\;10^{2}]\,, \label{scan}\\
T_R    &\in& [10^{4},\;f_a]\ {\rm GeV}\,, \nonumber
\eea
and take $m_{\tG} = 1$~TeV, $m_{\ta} = 100$~keV, $m_{\tz_1} = 150$~GeV and 
$\Omega_{\tz_1}^{std} h^2 = 10$, as before.
Our results will hardly depend on reasonable variation of these latter parameters, due to the enormous suppression
of long-lived relics due to saxion production and decay.
For each set of PQMSSM values the mis-alignment angle is chosen to enforce $\Omega_{DM} h^2 = 0.1123$.
The BBN bounds on late decaying saxions, neutralinos and gravitinos are once again applied and solutions
which satisfy all constraints are represented by blue dots. In order to differentiate the solutions
excluded due to late decaying neutralinos or gravitinos from solutions excluded due to late decaying saxions ($T_D < 5$~MeV),
we represent the former by red dots and the latter by green dots.

Fig.~\ref{fig:thetascan} shows the scan results for the misalignment angle $\theta_i$ versus $f_a$. 
As we can see, $f_a \sim 10^{16}$ requires $\theta_i \lesssim 0.07$. 
Although small $\theta_i$ values are still required in order to obtain $f_a \sim M_{GUT}$,
the mis-alignment angle can now be twenty times larger than in the non-SUSY PQ scenario
where the saxion/axino  fields are neglected. Furthermore, if we require
$f_a \sim 10^{15}$~GeV instead, $\theta_i$ can be as large as $0.4$. We also point
out that these conclusions are independent of our choice of $m_{\tz_1}$ and $\Omega_{\tz_1}^{std}$,
since the upper limit on $\theta_i$ comes entirely from the $T_D > 5$~MeV constraint.
These results also verify our estimate for $\theta_i^{max}$ in Eq.~(\ref{thetamax}),
which gives $\theta_i \lesssim 0.07-0.1 (0.4-0.6)$ for $f_a = 10^{16}(10^{15})$~GeV.

%
\FIGURE[t]{
\includegraphics[width=10cm]{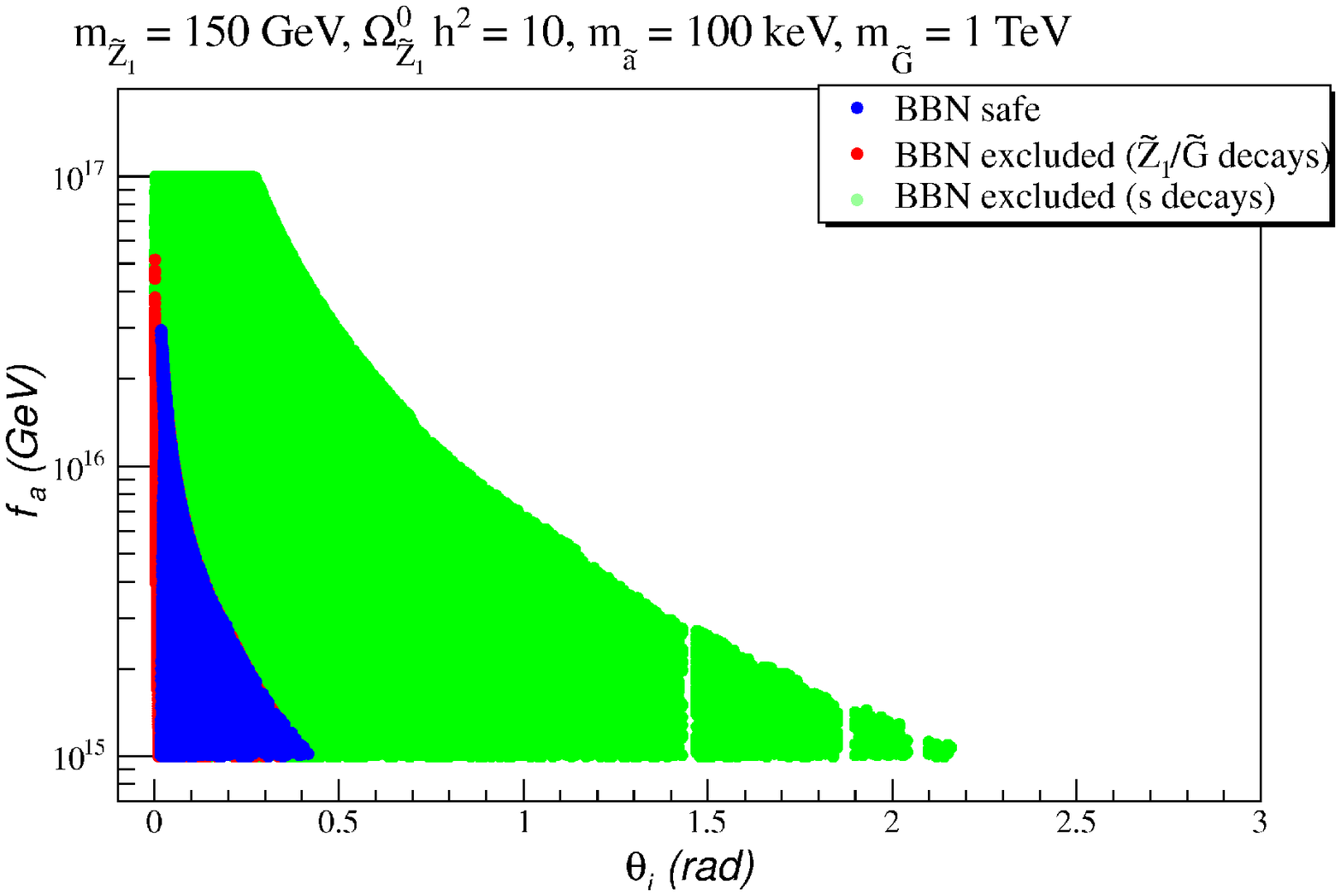}
\caption{The mis-alignment angle versus $f_a$ for a scan over the PQMSSM
parameters as discussed in the text. All points satisfy $\Omega_{DM} h^2 = 0.1123$,
while the green points are excluded by BBN constraints on late entropy injection from saxion
decays ($T_D < 5$~MeV). The red points are excluded by BBN bounds on late decaying neutralinos
or gravitinos and the blue points satisfy both the BBN and dark matter constraints.}
\label{fig:thetascan}}

In Fig.~\ref{fig:msscan}, we show the saxion mass versus $f_a$ for the
same points exhibited in Fig.~\ref{fig:thetascan}. As already
discussed in Sec.~\ref{sec:SD}, the BBN constraint on late decaying
saxions ($T_D > 5$~MeV) requires $m_{s}$ to be in the multi-TeV range, as shown by Eq.~(\ref{msmin}).
Since we expect $m_s\sim m_{3/2}$, models with large $m_{3/2}\sim 10-50$ TeV such as
Yukawa-unified SUSY\cite{yu}, mirage unification\cite{choi,cj}, effective SUSY\cite{ckn},
AMSB\cite{amsb} or string-motivated models such as G2-MSSM\cite{kane} would naturally yield such heavy saxions.
 
We also see that the BBN bounds on late decaying neutralinos do not
significantly constrain the saxion mass, since allowed (blue) solutions can
be found for any $m_s$ value as long as the bound in Eq.~(\ref{msmin})
is satisfied. We also point out that the few red points with 
saxion masses below the limit in Eq.~(\ref{msmin}) correspond to the case $T_e < T_D$,
where the saxions decay before dominating the energy density of the universe. All these
solutions have extremely small $\theta_i$ values, which lie in the narrow band
at $\theta_i \sim 0.001$ seen in Fig.~\ref{fig:thetascan}.

%
\FIGURE[t]{
\includegraphics[width=10cm]{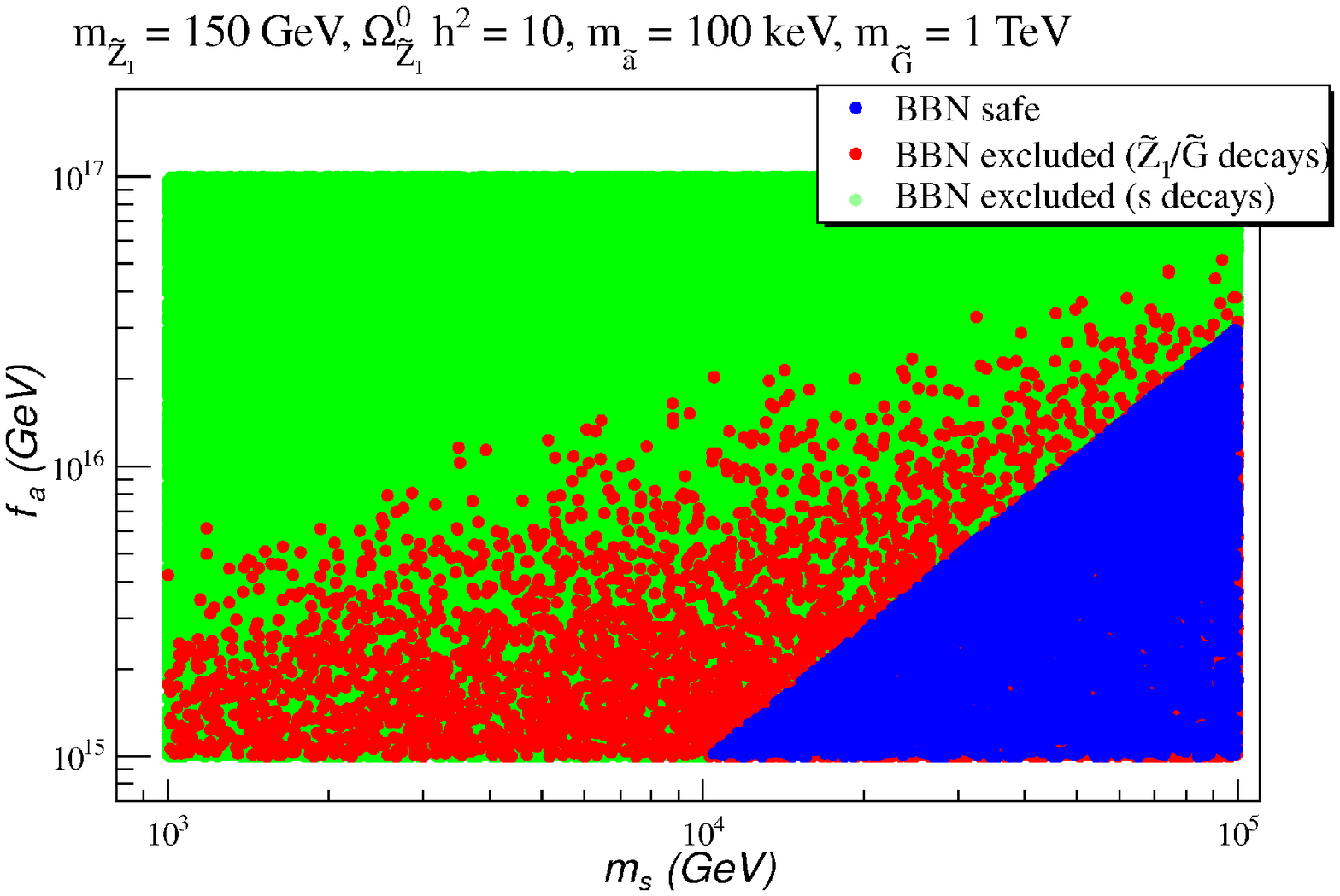}
\caption{The saxion mass versus $f_a$ for a scan over the PQMSSM
parameters as discussed in the text. All points satisfy $\Omega_{DM} h^2 = 0.1123$,
while the green points are excluded by BBN constraints on late entropy injection from saxion
decays ($T_D < 5$~MeV). The red points are excluded by BBN bounds on late decaying neutralinos
or gravitinos and the blue points satisfy both the BBN and dark matter constraints.}
\label{fig:msscan}}

Fig.~\ref{fig:siscan} shows the saxion field amplitude $s_i$ versus
$f_a$. As discussed in Sec.~\ref{sec:pheno}, $s_i$ parametrizes the
details of the transition from the static to the oscillatory regime of
the saxion field near $T=T_s$. To compute the value of $s_i$, the full saxion
potential for $T\gtrsim T_s$ needs to be known, which requires assuming
a specific PQMSSM model as well as knowledge of the SUSY breaking mechanism.
Nonetheless, natural values for $s_i$ are $f_a$ or $M_P$. Fig.~\ref{fig:siscan}
shows that small values of $s_i/f_a$ are disfavored, since
they suppress the entropy dilution of the neutralino and gravitino relic densities,
conflicting with the BBN bounds. However, as seen in Fig.~\ref{fig:siscan},
unnaturally large $s_i/f_a$ values are not necessary for
obtaining $f_a \sim M_{GUT}$.
Furthermore, smaller $\Omega_{\tz_1}^{std} h^2$ values would allow for smaller $s_i/f_a$.

%
\FIGURE[t]{
\includegraphics[width=10cm]{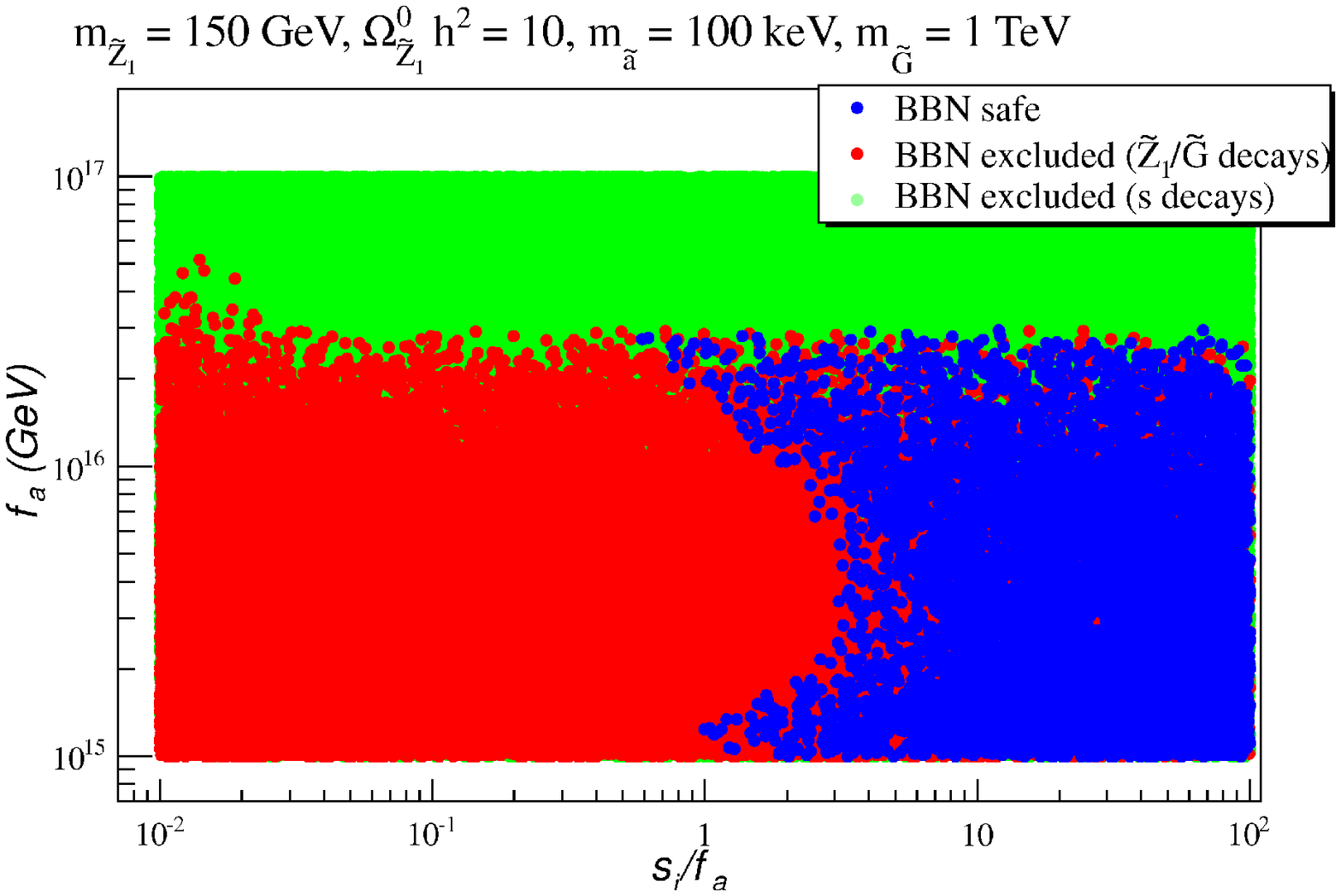}
\caption{The initial saxion field amplitude ($s_i$) divided by $f_a$ versus $f_a$ for a scan over the PQMSSM
parameters as discussed in the text. All points satisfy $\Omega_{DM} h^2 = 0.1123$,
while the green points are excluded by BBN constraints on late entropy injection from saxion
decays ($T_D < 5$~MeV). The red points are excluded by BBN bounds on late decaying neutralinos
or gravitinos and the blue points satisfy both the BBN and dark matter constraints.}
\label{fig:siscan}}

Finally-- in Fig.~\ref{fig:trscan}-- we show the reheat temperature versus $f_a$.
As already mentioned in Sec.~\ref{sec:SD}, large $T_R$ increases $T_e$,
resulting in an increase in the dilution of the axion, neutralino
and gravitino fields. From Fig.~\ref{fig:trscan}, we see that $T_R \gtrsim 10^8$~GeV
is usually required to satisfy the BBN constraints on late decaying neutralinos and gravitinos.
We point out that large reheat temperatures are motivated by thermal leptogenesis models\cite{bp,tr,ay}
which usually require  $T_R > T_R^{min} \sim 2\times 10^{9}$~GeV in order to explain the observed
matter-antimatter asymmetry ($\eta\equiv\frac{n_B-n_{\bar B}}{n_\gamma}\sim 6\times 10^{-10}$) in the 
universe\footnote{Here, $n_B$ is the baryon density, $n_{\bar B}$ is the antibaryon density and $n_\gamma$ is
the photon density of the universe.}.
However, due to the entropy injection during saxion decays, the asymmetry $n_B-n_{\bar B}$
will also be diluted by $1/r$. In this case the actual lower limit
on the reheat temperature is $T_R/r > T_R^{min}$. Hence, in frame {\it b} we show instead
$f_a\ vs.\ T_R/r$, and find that all BBN-allowed points have $T_R/r < 10^{6}$~GeV.
This rigid limit comes from the BBN bounds on late decaying gravitinos.

For $m_{\tG} \gtrsim M_i$ (as assumed here), where $M_i$ are the gaugino masses,
we have\cite{moroi}:
\be
\Omega_{\tG} h^2  \simeq  5\times 10^{-11} \left(\frac{m_{\tG}}{1\ {\rm TeV}}\right)\left(\frac{T_R}{r}\right)\;\; {\rm and}\;\;
\tau_{\tG} \simeq 10^{5}s \left(\frac{1\ {\rm TeV}}{m_{\tG}}\right)^{3} 
\label{grav}
\ee
where $\Omega_{\tG} h^2$ is the gravitino relic abundance including the entropy dilution.
Thus, for a 1 TeV gravitino, we have $\tau_{\tG} \simeq 10^5s$ and the BBN bounds described in Sec.~\ref{sec:bbn}
require $\Omega_{\tG} h^2 \lesssim 10^{-4}$, which implies $T_R/r \lesssim 10^{6}$~GeV, 
as seen in Fig.~\ref{fig:trscan}{\it b}.
Therefore, the unification of the PQ and GUT scales seems to strongly disfavor 
thermal leptogenesis\cite{leptog} scenarios,
unless a heavier gravitino is assumed, as in the usual MSSM scenario. In particular, for $m_{\tG} = 10$~TeV,
Eq.~(\ref{grav}) gives $\tau_{\tG} \simeq 10^2$s and the BBN bounds are considerably weaker in this case\cite{jedamzik}:
$\Omega_{\tG} h^2 \lesssim 0.5$. Hence, for multi-TeV gravitinos, we can have $T_R/r \lesssim 10^9$~GeV, which
makes thermal leptogenesis once again viable. 
We also note that non-thermal leptogenesis requires only $T_R/r \agt 10^6$ GeV\cite{ntlepto}, 
while Affleck-Dine leptogenesis allows still lower $T_R/r$ values\cite{affdine}.

%
\FIGURE[t]{
\includegraphics[width=10cm]{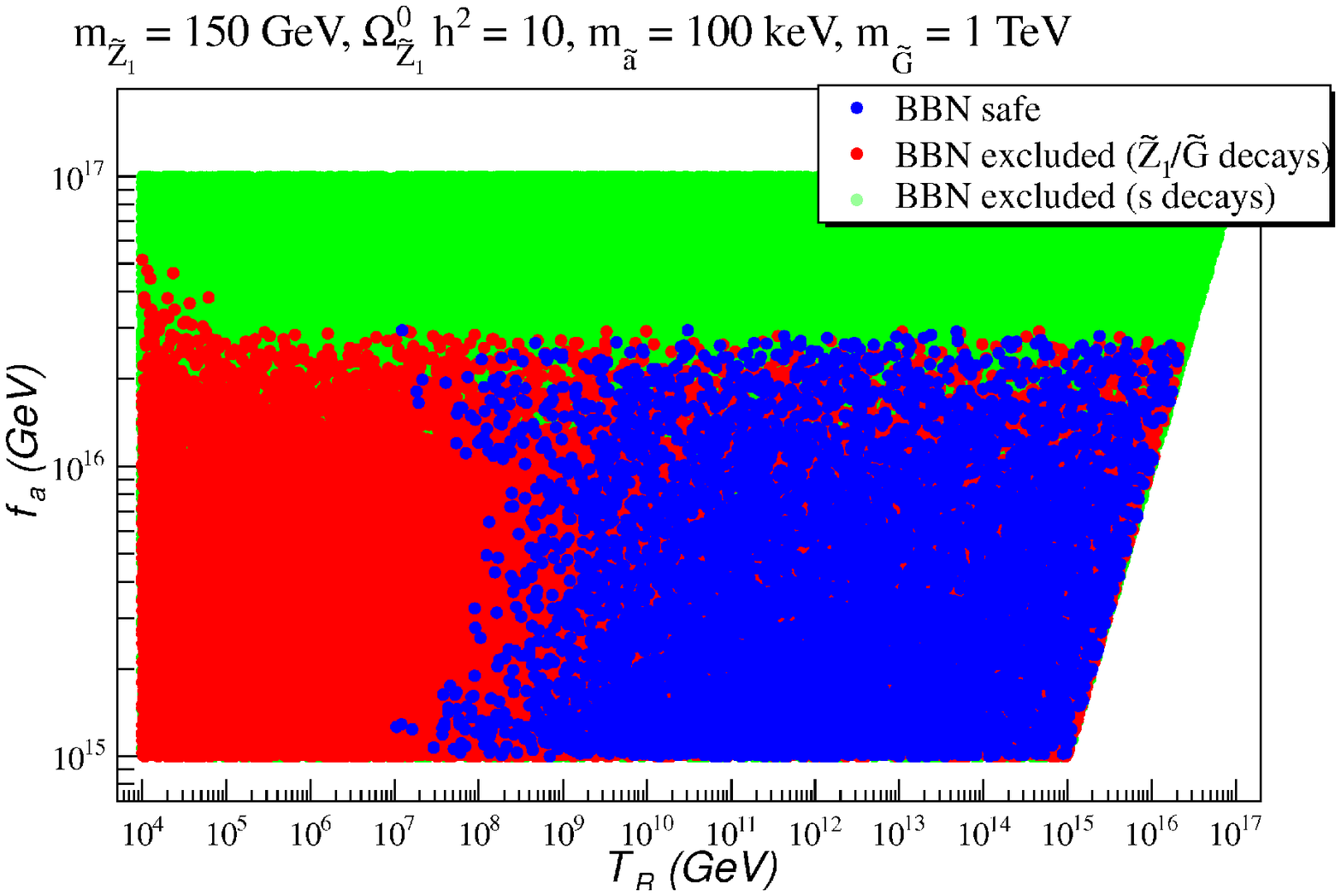}
\includegraphics[width=10cm]{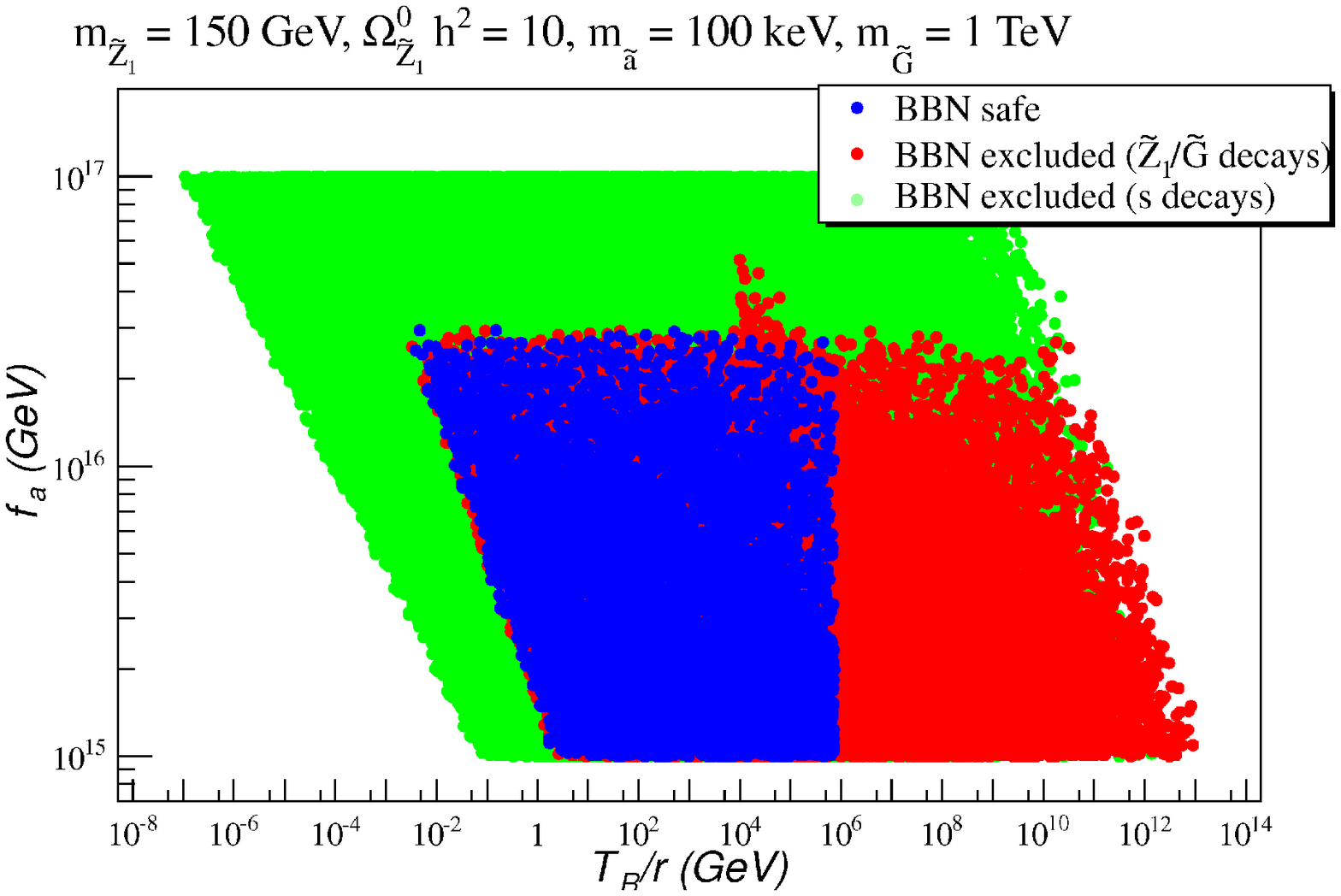}
\caption{In frame {\it a}, we plot the reheat temperature versus $f_a$ for a scan over the PQMSSM
parameters as discussed in the text. All points satisfy $\Omega_{DM} h^2 = 0.1123$,
while the green points are excluded by BBN constraints on late entropy injection from saxion
decays ($T_D < 5$~MeV). The red points are excluded by BBN bounds on late decaying neutralinos
or gravitinos and the blue points satisfy both the BBN and dark matter constraints.
In frame {\it b}, we show the value of $T_R/r$ to account for possible dilution of the
baryon asymmetry due to entropy production from saxions.
}
\label{fig:trscan}}

\section{The neutralino LSP case: mixed $a\tz_1$ DM}
\label{sec:az1}

So far, our results have focussed on the case of the PQMSSM with an axino as LSP, so 
that dark matter consists of an $a\ta$ mixture. 
Our results were largely independent of reasonable variations in $m_{\ta}$ since the
axino abundance suffers a huge suppression due both to the large values of $f_a$ which are required and to 
entropy production from saxion decays.
A qualitative difference results if we take $m_{\ta}$ so high that $m_{\ta}>m_{\tz_1}$ and
the neutralino $\tz_1$ becomes the LSP, so dark matter would consist of an $a\tz_1$ mixture.
In this case, gravitinos and axinos can still be produced thermally at high $T_R$, but now these
states will cascade decay down to the stable $\tz_1$ state, and possibly add to the thermal neutralino abundance.

In the mixed $a\tz_1$ DM scenario, neutralinos are produced via axino decay at 
temperature $T_D=\sqrt{\Gamma_{\ta}M_P}/(\pi^2g_*(T_D)/90)^{1/4}$, 
as well as via thermal freeze-out at $T=T_{fr}$. The neutralinos from axino decay
may {\it re-annihilate} at $T=T_D$ if the annihilation rate exceeds the expansion rate at $T_D$\cite{ckls}:
\be
\langle\sigma v\rangle \frac{n_{\tz_1}(T_D)}{s} > \frac{H(T_D)}{s}
\ee
where $n_{\tz_1}(T_D)$ is the total neutralino number density due to
thermal (freeze-out) and non-thermal (axino decays) production. 
Thus, the re-annihilation effect depends on a large thermal production rate for axinos.
In the $f_a\sim M_{GUT}$ case considered here, neutralino re-annihilation is largely irrelevant
because 1. thermal production of axinos is suppressed by $1/f_a^2$ and 2. 
the axino abundance at $T=T_D$ is also highly suppressed by the 
saxion entropy production.
Thus, for the large $f_a$ scenario, the neutralino abundance is estimated to be:
\be
\Omega_{\tz_1}=\Omega_{\tz_1}^{TP}+\frac{m_{\tz_1}}{m_{\ta}}\Omega_{\ta}^{TP}+
\frac{m_{\tz_1}}{m_{\tG}}\Omega_{\tG}^{TP} 
\ee
where $\Omega_{\tz_1}^{TP}$ is evaluated for either a MD, DD or RD universe, and $\Omega_{\ta}^{TP}$
and $\Omega_{\tG}^{TP}$ are diluted by entropy production ratio $r$ for $r>1$. 
At the end, we must add in the axion abundance $\Omega_a$ as calculated for a MD, DD or RD
universe, with the latter case diluted as usual by entropy ratio $r$ when $T_e<T_a$.

Our results in Sec. \ref{sec:axino} showed that models with a very light axino as LSP could
be consistent with $f_a\sim M_{GUT}$, but only with very large-- perhaps uncomfortably large-- 
values of the saxion mass, with $m_s$ typically in the tens of TeV range. 
In gravity-mediated SUSY breaking models, a puzzle would then arise as to why the sparticles
exist in the sub-TeV range, while saxions are present at 10-50 TeV.

Here, we note that there do exist several SUSY models where $m_s$ is naturally at the 
tens of TeV scale\cite{yu,ckn,amsb}.
One possibility consists of models with mixed moduli-anomaly mediated 
SUSY breaking soft terms (mirage unification, or MU)\cite{choi}.
These models are based on the KKLT proposal\cite{kklt} of string models where the moduli 
fields are stabilized via fluxes and the addition of a non-supersymmetric anti-D-brane breaks SUSY and 
provides an uplifting to the scalar potential leading to a deSitter vacuum.
In this class of models, Choi and Jeong have calculated the magnitude of soft SUSY 
breaking terms where the strong CP problem is solved by the PQ mechanism\cite{cj}. 
Since the MSSM soft terms arise from mixed moduli/anomaly mediation, their magnitude
is at the TeV scale even though $m_{3/2}$ is naturally in the multi-TeV regime. 
Furthermore, they find that typically $m_{\ta}\sim m_{\tG}$, and the saxion mass $m_s\sim \sqrt{2}m_{\tG}$.

Another possibility arises from $M$-theory models with compactification on manifolds of G2 holonomy. 
In these models\cite{kane}, 
the gravitino, axino and saxion masses, along with MSSM scalars, are all expected to inhabit the
multi-TeV range, while gaugino masses are expected to be much lighter, and non-universal. 
These latter models typically yield a wino-like lightest neutralino.

The MU soft terms have been programmed into Isajet/Isasugra\cite{isajet}, and are functions of 
the mixed moduli-AMSB mixing parameter $\alpha$, $m_{3/2}$, $\tan\beta$ and $sign(\mu )$.
They also depend on the matter and Higgs field modular weights $n_i$, which can take values of
$0$, $1/2$ or 1, depending on if the fields live on a D3-brane, a D7 brane or an intersection.
Spectra for the string models based on G2 holonomy can readily be generated with Isasugra using
non-universal gaugino mass entries. 

In Fig. \ref{fig:MM}{\it a})., we show the axion and neutralino relic abundances versus $f_a$ 
for the mirage-unification model with moduli/AMSB mixing parameter 
$\alpha =3$, $m_{3/2}=40$ TeV, $\tan\beta =10$ and $\mu >0$ with $m_t=173.3$ GeV. 
We take $m_{\ta}=m_{\tG}=40$ TeV, $m_s=\sqrt{2}m_{\tG}=56.6$ TeV and $T_R=10^{11}$ GeV. 
We further take
modular weights $n_m$ for matter fields equal to $1/2$ and for Higgs fields $n_H=1$.
The Isajet spectra gives $m_{\tg}=877.4$ GeV and $m_{\tz_1}=352$ GeV and $\Omega_{\tz_1}^{std}h^2=0.026$,
where $\tz_1$ is mainly higgsino-like. 
From the plot, we see that for $f_a\alt 1.5\times 10^{14}$ GeV, too much neutralino dark matter is
produced due to axino decays and already saturates the DM constraint, even for $\Omega_a h^2 = 0$.
As $f_a$ increases, the axino abundance falls sharply since the 
thermal production rate is suppressed by $1/f_a^2$ and the Yield is diluted by entropy injection
from saxion decays. The gravitino abundance also falls, but not as sharply, since here the
diminution is only due to entropy dilution from saxion decays. 
The thermal neutralino abundance falls, but less sharply still, 
since, for the choice of parameters in Fig.~\ref{fig:MM}, the neutralino freezes-out
in a MD universe, which results in a dilution smaller than $1/r$, as shown in Fig.~\ref{fig:omegas}. The relic 
axion abundance grows with $f_a$, but is also diluted. Here, we dial $\theta_i$ to an appropriate value
such that $\Omega_{a\tz_1}h^2$ is fixed at the mesured value of $\sim 0.11$. 
For $f_a \gtrsim 4\times 10^{14}$~GeV, the dark matter is axion-dominated. 
Once we reach values of $f_a\agt 10^{16}$ GeV, 
then the saxion decay temperature $T_D$ drops below 5 MeV and we consider the model BBN excluded (dashed curves).
When compared to Fig.~\ref{fig:OmegaR}, we see that the neutralino LSP case easily avoids the BBN bounds
on late decaying axinos, as already mention in Sec.~\ref{sec:bbn}.

In frame {\it b}.), we show the value of $\theta_i$ which is needed for the case of
$m_{3/2}=40$ and also for 30 TeV. The value of $\theta_i$ is again typically in the
$0.04-0.08$ range in order to suppress overproduction of axions.

The main results from the scan over parameter space performed for the axino LSP case,
shown in Fig's.~\ref{fig:thetascan}-\ref{fig:trscan}, still hold for the neutralino LSP scenario,
since they are weakly dependent on the nature of the LSP. 
However, since the BBN bounds on axino decays are easily avoided in the heavy axino case,
the lower bounds on $T_R$ and $s_i/f_a$ are now relaxed in the neutralino LSP scenario.
Nonetheless, the upper bound on $T_R/r$, relevant for baryogenesis mechanisms, still
holds (if we keep $m_{\tG}$ = 1 TeV), since it only depends on the gravitino mass,
as discussed in the last Section.

%
\FIGURE[t]{
\includegraphics[width=10cm]{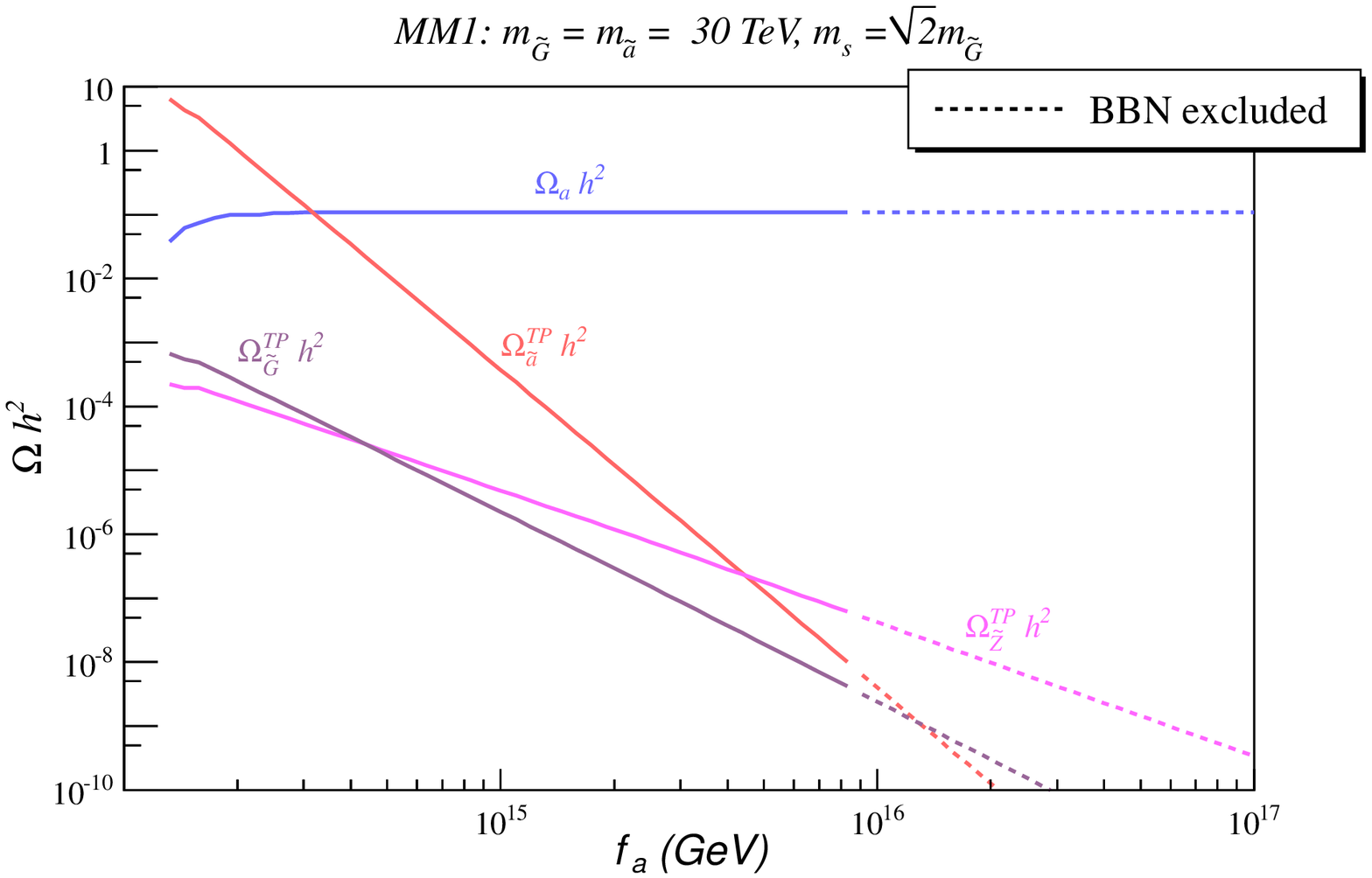}
\includegraphics[width=10cm]{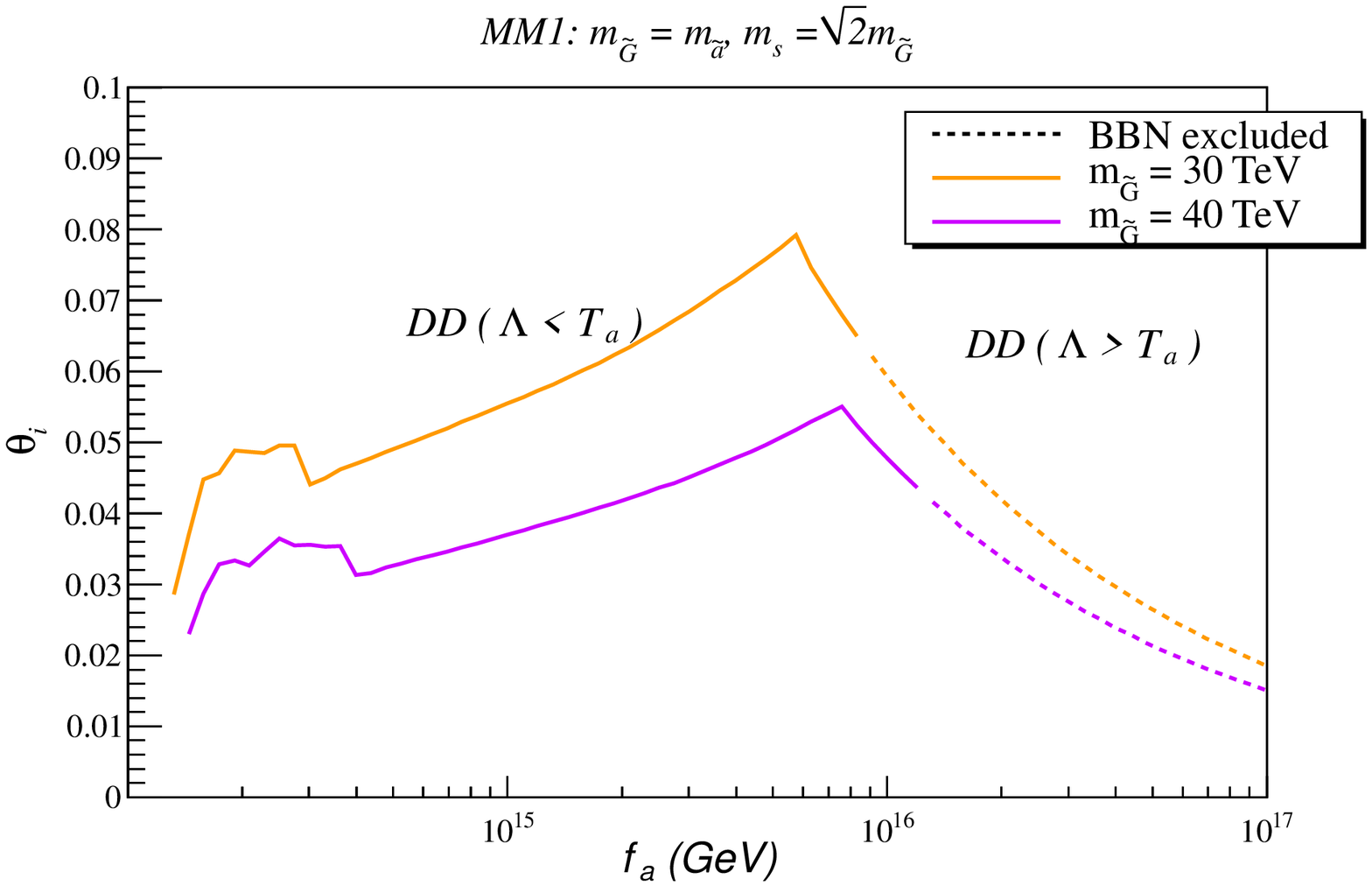}
\caption{The axion, axino, neutralino and gravitino relic densities as a function of $f_a$ for
the Mirage Unification model of SUSY breaking, with 
$\alpha =3$, $m_{3/2} = 40$~TeV, $\tan\beta =10$ and $\mu >0$. 
We also take $m_{\ta} = m_{3/2}$ and $m_{s} =\sqrt{2}m_{3/2}$, with $T_R=10^{11}$ GeV and $s_i = 10f_a$. 
The misalignment angle ($\theta_i$) is chosen such as 
$\Omega_{a\tz_1} h^2 \simeq \Omega_{a} h^2 = 0.1123$.
The dashed region is excluded by BBN bounds on saxion decay: $T_{D} < 5$~MeV.
In frame {\it b})., we show the required value of misalignment angle $\theta_i$ for the 
case of $m_{3/2}=40$ TeV and also for $m_{3/2}=30$ TeV.
}
\label{fig:MM}}

\section{Conclusions}
\label{sec:conclude}

We have investigated the possibility of extending the usual upper bound on
the Peccei-Quinn scale ($f_a$) to values of the order of the Grand Unification scale ($M_{GUT}$).
We show that large $f_a$ values usually lead to an early universe dominated by coherent oscillating
saxions, which are required to decay before Big Bang nucleosynthesis. 
In the case of a light axino as LSP, the injection of entropy during
the decay of the saxion field results in a dilution of the axion relic density,
which allows us to evade the usual upper bound on $f_a$ ($\lesssim 10^{12}$~GeV).
Furthermore, the dilution of the neutralino and gravitino relic densities naturally
evade the BBN bounds on late decaying $\tz_1$s and $\tG$s. From Eq's. (\ref{thetamax}) and
(\ref{msmin}), verified by the scan over parameter space, we find that, in order to allow $f_a \sim 10^{16}$ GeV:
\bi
\item $\theta_i \alt 0.07-0.1$ is necessary to satisfy the axionic dark matter relic density constraint,
\item $m_s \agt 50$~TeV in order to satisfy BBN constraints on late decaying saxions.
\ei
Furthermore, for the axino LSP case with a neutralino LSP with $m_{\tz_1} = 150$~GeV and $\Omega_{\tz_1} h^2 = 10$:
\bi
\item $s_i/f_a \gtrsim 1$ and $T_R \gtrsim 10^8$~GeV are required to increase coherent saxion 
production and hence increase  dilution of the neutralino relic densities 
and to satisfy the BBN bounds.
\ei
While the first two conditions are quite independent of the SUSY spectrum chosen (parametrized here
by $\Omega_{\tz_1}^{std}h^2$ and $m_{\tz_1}$), the third condition can be relaxed
if PQMSSM models with heavier neutralinos and/or smaller $\Omega_{\tz_1}^{std}$ are considered.

We also investigated the case where $m_{\ta}\sim m_{3/2}$ and the neutralino is LSP such that
dark matter is comprised of an axion/neutralino mixture. Models such as Mirage Unification or
string models based on G2 holonomy naturally give axino and saxion masses in the tens of TeV range,
while maintaining at least some superpartners below the TeV scale. 
In these models, if $f_a\sim M_{GUT}$, then again we expect large amounts of entropy production from
saxion decay, while neutralino, axino and gravitino abundances are all supressed to tiny levels, 
thus helping to avoid BBN constraints. 

While all our results were caluclated assuming the saxion $s\to gg$ decay mode at 100\%, we note that
other model-dependent decay modes such as $s\to hh$ or $s\to aa$ may be present. The first of these
would contribute to additional entropy production and decrease the saxion lifetime, 
thus helping to avoid BBN constraints. In this sense, we regard our results as conservative.
On the other hand, if saxion decay into axions is significant, then saxion decays inject relativistic axions
and increase the effective value of $g_*$ during BBN. 
In addition, the entropy injection from saxion production and decay is greatly diminished.
In this case, in order to keep the sucessful BBN predictions,
$\rho_s$ has to be suppressed\cite{kns}, which likely disfavors the $f_a \sim M_{GUT}$ scenario.

As consequences of the $f_a\sim M_{GUT}$ scenario, we would expect the DM of the universe
to be axion-dominated, with a tiny component of either axinos or neutralinos.
The axion mass is expected to lie in the $10^{-10}$ eV range which is well below the range
currently being explored by the ADMX experiment\cite{admx}. New ideas 
and new experiments will likely be needed to explore the axion direct detection signal
in this mass range. Furthermore, the $f_a \sim M_{GUT}$ case can accommodate a much
wider range of $\Omega_{\tz_1} h^2$ values than the pure neutralino DM scenario in MSSM models,
since, for a neutralino LSP, its relic abundance is suppressed by the entropy injection,
while, for an axino LSP, the neutralino contribution to the DM relic abundance is suppressed both by
entropy dilution and $m_{\ta}/m_{\tz_1}$.

{\it Note added:} As this manuscript was nearing completion, a paper by
Kawasaki {\it et al.} Ref. \cite{kkn} was released; they also show that large $f_a\sim M_{GUT}$ can occur
in an inflationary model which relates inflation to the PQ scale.

\acknowledgments

This research was supported in part by the U.S. Department of Energy,
by the Fulbright Program and CAPES (Brazilian Federal Agency for
Post-Graduate Education).

\appendix
\section{Relic Densities in a Saxion Dominated Universe}

Here we briefly review the cosmology of an early saxion dominated universe and present
the expressions for the axion and neutralino relic densities used in Sec.~\ref{sec:bigfa}.

As discussed in Sec.~\ref{sec:SD}, we assume that the saxion field becomes the
main component of the universe's energy density at a temperature $T_e$ given by
Eq.~(\ref{TE}). Therefore, the universe becomes matter dominated until the saxion
decays at $T=T_D$, with the decay temperature given by Eq.~(\ref{TD}).

During the saxion dominated phase ($T_D < T < T_e$), the universe expands at a faster rate, given by:
\be
H(T) = \frac{1}{M_P} \sqrt{\frac{\rho_s}{3}} = \sqrt{\frac{\pi^2}{90} g_*(T_e)}\frac{T_e^2}{M_P}\left(\frac{R(T_e)}{R(T)}\right)^{3/2}
\ee
where $g_*(T)$ counts the number of relativistic degrees of freedom\footnote{For simplicity, we assume $g_*(T) = g_{*S}(T)$, where
$\rho_{rad} = \pi^2 g_*(T) T^4/30$ is the radiation energy density and $s = 2 \pi^2 g_{*S}(T) T^3/45$ is the entropy density of
the universe.}, $R(T)$ is the scale factor, $M_{P}$ is the reduced Planck Mass and
we have used Eq.~(\ref{TE}). For $T \gg T_D$, most of the saxions have not yet decayed and entropy is conserved:
\be
\frac{g_*(T) T^3}{g_*(T_e) T_e^3} = \left(\frac{R(T_e)}{R(T)}\right)^3 .
\ee
Therefore,
\be
H(T) = \frac{1}{3} \sqrt{\frac{ \pi^2}{10} g_*(T) T_e} \frac{T^{3/2}}{M_P} \; \mbox{ ($T_D \ll T < T_e$ )} . 
\label{HMD}
\ee

On the other hand, if $T\gtrsim T_D$, the entropy injection of saxions results in\cite{entropy}:
\begin{eqnarray*}
\frac{g_*(T)^2 T^8}{g_*(T_D)^2 T_D^8} = \left(\frac{R(T_D)}{R(T)}\right)^3  \\
\To H(T) = H(T_D) \frac{g_*(T) T^4}{g_*(T_D) T_D^4} .
\end{eqnarray*}
Since at $T=T_D$ the energy density is once again radiation dominated, we have:
\begin{eqnarray}
H(T_D) & = & \sqrt{\frac{\pi^2}{90} g_*(_D)}\frac{T_D^2}{M_P} \nonumber \\
\To H(T) & = & \sqrt{\frac{\pi^2}{90}}\frac{g_*(T)}{\sqrt{g_*(T_D)}}\frac{T^4}{T_D^2 M_P}  \; \mbox{ ($T_D < T \ll T_e$ )} \label{HDD}
\end{eqnarray}

\subsection{The Axion Relic Density}

Due to the distinct expressions for $H(T)$ during the matter dominated (MD)
or decaying particle dominated (DD) phases,
the axion oscillation temperature ($T_a$) as well as the coherent axion oscillation relic density
can significantly deviate from the usual expressions, if the axion starts to
oscillate in the saxion dominated era ($T_D < T_a < T_e$). Using Eqs.~(\ref{HMD}), (\ref{HDD}) and Eq.~(\ref{ma})
for the $T$ dependent axion mass and Eq.~(\ref{oscond}) to define $T_a$ we find:
\be
\Omega_a = \left\{ \begin{array}{ll}
\Omega_a^{RD}/r \;  \mbox{, if $T_e < T_a$} \\
\Omega_a^{MD}\;  \mbox{, if $T_{DD} < T_a < T_e$} \\
\Omega_a^{DD}\;  \mbox{, if $T_D < T_a < T_{DD}$} \\
\Omega_a^{RD}\;  \mbox{, if $T_a < T_D$}
\end{array} \right.
\ee
where $r$ is defined in Eq.~(\ref{r}) and\footnote{All quantities are in GeV units.}
\be
\Omega_a^{RD} h^2 = \left\{ \begin{array}{ll}
9.23\times10^{-3} \theta_i^2 f(\theta_i)  \frac{1}{g_*(T_a)^{1/4}} \left(\frac{f_a}{10^{12}}\right)^{3/2} \mbox{, if $T_a < \Lambda$} \\
1.32\ \theta_i^2 f(\theta_i) \frac{1}{g_*(T_a)^{5/12}}\left(\frac{f_a}{10^{12}}\right)^{7/6} \mbox{, if $T_a > \Lambda$} \label{case0}
\end{array} \right.
\ee
\be
\Omega_a^{MD} h^2 = \left\{ \begin{array}{ll}
 7.5\times10^{-5}\ \theta_i^2 f(\theta_i) T_D \left(\frac{f_a}{10^{12}}\right)^2 \mbox{, if $T_a < \Lambda$} \\
 1.4\ \theta_i^2 f(\theta_i) \frac{1}{g_*(T_a)^{4/11}}\left(\frac{f_a}{10^{12}}\right)^{14/11} \frac{T_D}{T_e^{4/11}} \mbox{, if $T_a > \Lambda$}
\end{array} \right.
\ee
\be
\Omega_a^{DD} h^2 = \left\{ \begin{array}{ll}
 7.5\times10^{-5}\ \theta_i^2 f(\theta_i) T_D \left(\frac{f_a}{10^{12}}\right)^2 \mbox{, if $T_a < \Lambda$} \\
 1.72\ \theta_i^2 f(\theta_i) \frac{g_*(T_D)^{1/4}}{\sqrt{g_*(T_a)}} T_D^2 \left(\frac{f_a}{10^{12}}\right)^{3/2} \mbox{, if $T_a > \Lambda$} 
\end{array} \right. .
\ee
The corresponding expressions for the oscillation temperature are:
\be
T_a^{RD} = \left\{ \begin{array}{ll}
1.23\times 10^2  \frac{1}{g_*(T_a)^{1/4}} \left(\frac{10^{12}}{f_a}\right)^{1/2} \mbox{, if $T_a^{RD} < \Lambda$} \\
8.71\times10^{-1} \frac{1}{g_*(T_a)^{1/12}} \left(\frac{10^{12}}{f_a}\right)^{1/6} \mbox{, if $T_a^{RD} > \Lambda$}
\end{array} \right.
\ee
\be
T_a^{MD} = \left\{ \begin{array}{ll}
6.1\times10^2 \left(\frac{1}{\sqrt{g_*(T_a) T_e}} \frac{10^{12}}{f_a} \right)^{2/3}  \mbox{, if $T_a^{MD} < \Lambda$} \\
8.6\times10^{-1} \left(\frac{1}{\sqrt{g_*(T_a) T_e}} \frac{10^{12}}{f_a} \right)^{2/11}  \mbox{, if $T_a^{MD} > \Lambda$}
\end{array} \right. 
\ee
\be
T_a^{DD} = \left\{ \begin{array}{ll}
0.11\times10^2 \left(\frac{\sqrt{g_*(T_D)}}{g_*(T_a)} \frac{10^{12}}{f_a} T_D^2 \right)^{1/4}  \mbox{, if $T_a^{DD} < \Lambda$} \\
9.0\times10^{-1} \left(\frac{\sqrt{g_*(T_D)}}{g_*(T_a)} \frac{10^{12}}{f_a} T_D^2 \right)^{1/8} \mbox{, if $T_a^{DD} > \Lambda$}
\end{array} \right. .
\ee
The temperature $T_{DD}$ marks the transition from the matter dominated phase to the decaying particle dominated phase,
where entropy is no longer conserved. An approximate value for $T_{DD}$ can be obtained matching the axion relic densities in
the MD and DD phases:
\be
T_{DD} = \left(\frac{g_*(T_D)}{g_*(T_a)} T_e T_D^4\right)^{1/5}
\ee

\subsection{The Neutralino Relic Density}

The neutralino will decouple from the thermal bath when
\be
\langle \sigma v\rangle n_{\tz_1}(T_{fr}) = H(T_{fr}) ,\label{Tfr}
\ee
where $T_{fr}$ is the freeze-out temperature, $\langle \sigma v\rangle$ is the neutralino annihilation cross-section and
\be
n_{\tz_1}(T) = 2 \left(\frac{m_{\tz_1} T}{2 \pi}\right)^{3/2} e^{-m_{\tz_1}/T} .
\ee
In a radiation dominated universe, the neutralino yield is given by:
\be
Y_z(T_{fr}) = \frac{H(T_{fr})}{\langle \sigma v\rangle s(T_{fr})}  ,
\label{yRD}
\ee
while in a matter dominated universe
\be
Y_z(T_{fr}) = \frac{3}{2} \frac{H(T_{fr})}{\langle \sigma v\rangle s(T_{fr})} .
\label{yMD}
\ee

As in the axion case, the neutralino can freeze-out before the universe becomes matter dominated ($T_{fr} > T_e$),
during the matter dominated phase ($T_D \ll T_{fr} \lesssim T_e$),
during the decay dominated phase ($T_D \lesssim T_{fr} \ll T_e$) or during the radiation dominated phase ($T_{fr} < T_D$).
Using Eqs.~(\ref{HMD}), (\ref{HDD}) and (\ref{Tfr})-(\ref{yMD}), we obtain:

\be
\Omega_{\tz_1} = \left\{ \begin{array}{ll}
\Omega_{\tz_1}^{RD}/r \;  \mbox{, if $T_e < T_{fr}$} \\
\Omega_{\tz_1}^{MD}\;  \mbox{, if $T_{DD} < T_{fr} < T_e$} \\
\Omega_{\tz_1}^{DD}\;  \mbox{, if $T_D < T_{fr} < T_{DD}$} \\
\Omega_{\tz_1}^{RD}\;  \mbox{, if $T_{fr} < T_D$}
\end{array} \right.
\ee
where
\bea
\Omega_{\tz_1}^{RD} h^2 &=& 8.5\times10^{-11} \frac{1}{\sqrt{g_*(T_{fr})}} \left(\frac{m_{\tz_1}}{T_{fr}}\right) \left( \frac{{\rm GeV^{-2}}}{\langle \sigma v\rangle}\right) \nonumber \\
\Omega_{\tz_1}^{MD} h^2 &=& \frac{3}{2} \frac{T_D}{\sqrt{T_e T_{fr}}} \times \Omega_{\tz_1}^{RD} h^2\\
\Omega_{\tz_1}^{DD} h^2 &=& \frac{3}{2} \sqrt{\frac{g_*(T_D)}{g_*(T_{fr})}} \frac{T_D^3}{T_{fr}^3} \times \Omega_{\tz_1}^{RD} h^2 \nonumber .
\eea
The corresponding freeze-out temperatures are given by:

\bea
T_{fr}^{RD} = m_{\tz_1}/\ln[\frac{3\sqrt{5}}{\pi^{5/2}} \langle \sigma v\rangle M_P m_{\tz_1}^{3/2} \frac{1}{\sqrt{g_*(T_{fr}) T_{fr}}}] \label{tfr0} \nonumber \\
T_{fr}^{MD} = m_{\tz_1}/\ln[\frac{3\sqrt{5}}{\pi^{5/2}} \langle \sigma v\rangle M_P m_{\tz_1}^{3/2} \frac{1}{\sqrt{g_*(T_{fr}) T_e}}]  \\
T_{fr}^{DD} = m_{\tz_1}/\ln[\frac{3\sqrt{5}}{\pi^{5/2}} \langle \sigma v\rangle M_P m_{\tz_1}^{3/2} \frac{\sqrt{g_*(T_D)}}{g_*(T_{fr})}\frac{T_D^2}{T_{fr}^{5/2}}] \nonumber
\eea
Once again $T_{DD}$ marks the transition between the MD and DD phases and is given by:
\be
T_{DD} = \left(\frac{g_*(T_D)}{g_*(T_{fr})} T_e T_D^4\right)^{1/5} .
\ee

%

\end{document}